 \documentclass[12pt,preprint]{aastex}

\begin{document}

\title{Gamma-Ray Bursts and the Earth: 
	Exploration of Atmospheric, Biological, Climatic and 
	Biogeochemical Effects}



\author{Brian C. Thomas\altaffilmark{1,5}, Adrian L. Melott\altaffilmark{1},
	Charles H. Jackman\altaffilmark{2}, Claude M. Laird\altaffilmark{1,4},
	Mikhail V. Medvedev\altaffilmark{1},
	Richard S. Stolarski\altaffilmark{2}, 
	Neil Gehrels\altaffilmark{3}, John K. Cannizzo\altaffilmark{3},
	Daniel P. Hogan\altaffilmark{1}, and Larissa M. Ejzak\altaffilmark{1}
	}		

\altaffiltext{1}{University of Kansas, Department of Physics and Astronomy,
		 Lawrence, KS  66045-7582; bthomas@ku.edu, melott@ku.edu, claird@ku.edu,
		 dphogan@ku.edu, cnfsdlma@ku.edu}
\altaffiltext{2}{Laboratory for Atmospheres,
		 NASA Goddard Space Flight Center,
		 Code 613.3,	
		 Greenbelt, MD 20771; jackman@assess.gsfc.nasa.gov,
		 stolar@polska.gsfc.nasa.gov}
\altaffiltext{3}{Laboratory for Astroparticle Physics,
	  	 NASA Goddard Space Flight Center,
		 Code 661,
		 Greenbelt, MD 20771; gehrels@milkyway.gsfc.nasa.gov, 
		 cannizzo@milkyway.gsfc.nasa.gov}
\altaffiltext{4}{Also at Haskell Indian Nations University}
\altaffiltext{5}{After August, 2005, at Washburn University, Department of Physics and Astronomy,
		 Topeka, KS 66621; brian.thomas@washburn.edu}

\begin{abstract} 
Gamma-Ray Bursts (GRBs) are likely to have made a number of significant impacts on the Earth during the last billion years. The gamma radiation from a burst within a few kpc would quickly deplete much of the Earth's protective ozone layer, allowing an increase in solar UVB radiation reaching the surface.  This radiation is harmful to life, damaging DNA and causing sunburn.  In addition, $\mathrm{NO_2}$ produced in the atmosphere would cause a decrease in visible sunlight reaching the surface and could cause global cooling.  Nitric acid rain could stress portions of the biosphere, but the increased nitrate deposition could be helpful to land plants. 

We have used a two-dimensional atmospheric model to investigate the effects on the Earth's atmosphere of GRBs delivering a range of fluences, at various latitudes, at the equinoxes and solstices, and at different times of day.  We have estimated DNA damage levels caused by increased solar UVB radiation, reduction in solar visible light due to $\mathrm{NO_2}$ opacity, and deposition of nitrates through rainout of $\mathrm{HNO_3}$.
For the ``typical'' nearest burst in the last billion years, we find globally averaged ozone depletion up to 38\%.  Localized depletion reaches as much as 74\%.  Significant global depletion (at least 10\%) persists up to about 7 years after the burst.  Our results depend strongly on time of year and latitude over which the burst occurs.  The impact scales with the total fluence of the GRB at the Earth, but is insensitive to the time of day of the burst and its duration (1s to 1000s).   We find DNA damage of up to 16 times the normal annual global average, well above lethal levels for simple life forms such as phytoplankton.  The greatest damage occurs at low to mid latitudes.  We find reductions in visible sunlight of a few percent, primarily in the polar regions.  Nitrate deposition similar to or slightly greater than that currently caused by lightning is also observed, lasting several years.
We discuss how these results support the hypothesis that the Late Ordovician mass extinction may have been initiated by a GRB.
\end{abstract}


\section{Introduction}
Knowledge of the effects of gamma-ray bursts (GRBs) on the Earth and life is important for establishing where and when life is likely to exist in our own Galaxy and others.  Concepts such as the Galactic Habitable Zone \citep{gonz01} depend on information about the rate and distance of effectiveness of GRBs, along with other transient radiation events such as supernovae.  In addition, GRBs may have played a role in mass extinctions in the Earth's history (e$.$g$.$ \citet{mel04} and references therein).

As discussed in \citet{mel04} and \citet{der05} it is likely that a GRB has occurred at least once in the last Gy close enough to have dramatic effects on stratospheric ozone, leading to detrimental effects on life through increases in solar ultraviolet (UV) radiation which is strongly absorbed by ozone.

In \citet{thom05} we presented results of an initial set of parameters modeling the impact of a gamma-ray burst on the earth's atmosphere and subsequent effects on life.  
In this study, we present further results for various parameter values.  In particular, we vary incident fluence, time of year of the burst, impact latitude, time of day, and duration of the burst.  


\section{Methods}  
\subsection{Atmospheric Model}
Our modeling of the atmospheric effects of a GRB was performed using the Goddard Space Flight Center two-dimensional atmospheric model, which was also used in the study reported in \citet{thom05}. We briefly describe the model here, and give a more detailed description in Appendix~\ref{app:model}.

The model's two spatial dimensions are altitude and latitude.  The latitude range is divided into 18 equal bands and extends from pole to pole.  The altitude range includes 58 evenly spaced logarithmic pressure levels (approximately 2 km spacing) from the ground to approximately 116 km.  A lookup table is used for computation of photolytic source term, used in calculations of photodissociation rates of atmospheric constituents by sunlight \citep{jack96}.  Winds and small scale mixing are included as described in \citet{flem99} (see Sect.~\ref{model-detail-trans}).  

We have employed two versions of the atmospheric model.  One is intended for long term runs (many years) and includes all transport mechanisms (e.g., winds and diffusion); it has a time step of one day and computes daily averaged constituent values.  The second is used for short term runs (a few days) and calculates constituent values throughout the day and night, but does not include full transport.  Previously, this version has been used with a time step of 225 seconds \citep{jack01}.  In the current study we have used a time step of one second in order to allow for inputting our GRB ionization over several time steps and to properly account for chemistry during and after the burst (see also Sect.~\ref{model-detail-chem}).

We have not included the effects of any ultra high energy ($>10^{18}~\mathrm{eV}$) cosmic rays from a GRB \citep{der04,wax04a,wax04b,der05}, due to uncertainty as to whether and at what energies GRBs may produce such particles. 

\subsection{GRB Ionization Rate Profiles}
\label{model-grbinput}
Though there is much variation among GRBs, estimates of the typical isotropic equivalent power of low redshift GRBs range from $4.4\times 10^{44} ~\mathrm{W}$ \citep{gue05} to $6\times 10^{44} ~\mathrm{W}$ \citep{llo02}.  Therefore, for the purposes of this study, we assume a ``standard'' burst with power $5\times10^{44}~\mathrm{W}$ (isotropic equivalent).  Burst durations (considering only ``long'' GRBs) range from $1.66$ to $278~\mathrm{s}$ \citep{preece00}.  We assume for this study a standard burst duration of  $10~ \mathrm{s}$.  We also investigate bursts of 1, 100 and 1000 s duration.  We do not consider short bursts due the lack of data on rate, luminosity, distance, etc.

In \citet{mel04} we estimated conservatively that the probable nearest burst pointed at the Earth in the last billion years is approximately 2 kpc.  For our standard burst this corresponds to a fluence of $100~\mathrm{kJ/m^2}$.  \citet{der05} estimated this distance to be 1 kpc.  
Our estimate of the nearest probable burst distance in the last billion years is arrived at by scaling from the universal observed rate (as of 2003), taking into account the universal star formation rate, and using the blue luminosity surface density of the Galaxy.  We then use the fluence from our standard burst at this distance to compute atmospheric and other effects.

Ionization rate profiles due to a gamma-ray burst are input to the model as production sources of $\mathrm{NO_y}$ and $\mathrm{HO_x}$.  It is assumed that for each ion-electron pair produced, 1.25 $\mathrm{NO_y}$ molecules are produced at all pressure levels \citep{pjg76} and 2.0 molecules of $\mathrm{HO_x}$ are produced below 75 km and less than 2.0 (from 1.99 to 0.0) for altitudes greater than 75 km \citep{sol81}.

The ionization rate profiles are calculated separately from the atmospheric model, following the method used in \citet{geh03}.  This section draws on that publication as well as an unpublished manuscript by Claude Laird.  In the \citet{geh03} study, gamma-rays were included using the spectrum of SN 1987A.  In the present study, the gamma-ray differential photon count spectrum used is that of \citet{band93}, which consists of two smoothly connected power laws:
\begin{eqnarray}
{dN(E) \over dE} & = & A\left({E \over 100~\mathrm{keV}}\right)^{\alpha} \exp{\left({-E \over E_0}\right)},  \nonumber \\
                 & ~ & E \leq (\alpha - \beta)E_0 ~, \nonumber \\
                  \\
                 & = &A\left[{(\alpha - \beta)E_0 \over 100~\mathrm{keV}}\right]^{\alpha - \beta} \exp(\beta - \alpha)\left({E \over 100~\mathrm{keV}}\right)^{\beta}, \nonumber \\
                 & ~ & E \geq (\alpha - \beta)E_0 ~,\nonumber
\end{eqnarray}
where we take typical values $E_0 = 187.5~ \mathrm{keV}$, $\alpha = -0.8$, $\beta = -2.3$ \citep{preece00}.  We have not taken variability of these quantities during the burst into account.
The flux constant, $A$, is used to scale the total received energy to our desired value (in this study, corresponding to fluences of $10~ \mathrm{kJ/m}^2$, $100~ \mathrm{kJ/m}^2$ and $1~ \mathrm{MJ/m}^2$).  

The total photon flux in each of 66 evenly spaced logarithmic energy bins, ranging $0.001 \leq E \leq 10 ~ \mathrm{MeV}$, is obtained by integrating the above spectrum for each bin.
The high and low energy regions of the spectrum are integrated separately; the $E \geq (\alpha - \beta)E_0$ region is integrated analytically and the $E \leq (\alpha - \beta)E_0$ energy region is numerically integrated using the ``qsimp'' routine from Numerical Recipes \citep{nr}.

The total incident energy flux in a monoenergetic beam at the top of the atmosphere is $F_i{^0} = N_i{^0}\langle E_i \rangle$, where $N_i{^0}$ is the incident photon flux (calculated from the Band spectrum as described above).
The photon flux is propagated vertically through a standard atmosphere (adjusted for the appropriate latitude and time when input to the atmospheric model), and is attenuated with altitude by an exponential decay law, with frequency-dependent absorption coefficients taken from a lookup table \citep{ple81}.  

The photon flux for the $i$th energy bin which remains at the $j$th altitude layer is given by 
$N_{i,j} = N_i{^0}e^{-\mu_i x_j}$, 
where $x_j$ is the column density (in $\mathrm{g/cm^2}$ measured from the top of the atmosphere and $\mu_i$ is the absorption coefficient 
(from \citet{ple81}).
The photon flux deposited in the $j$th layer with energy $\langle E_i \rangle$ is $\Delta N_{i,j} = N_{i,j-1} - N_{i,j}$.  The corresponding energy flux deposited is $F_{i,j} = \Delta N_{i,j}\langle E_i \rangle$ (in $\mathrm{MeV/cm^2/s}$).  The total ionization rate is the sum over all energies:
\begin{equation}
q_{tot,j} = {1 \over 35 ~\mathrm{eV}} \sum_{i=1}^{66} {F_{i,j} \over \Delta Z_j}
\end{equation}
where 35 eV is the energy needed to produce one ion pair \citep{pjg76} and $\Delta Z_j$ is the thickness of the $j$th altitude layer.

The vertical ionization rate profiles are mapped onto an altitude-latitude grid for use by the atmospheric model as follows.  First, a series of ionization rate profiles as functions of altitude and latitude are calculated for a flat Earth for incidence angles of $5^\circ$ -- $85^\circ$ from zenith, corresponding to latitude bands $10^\circ$ wide centered on $85^\circ$ -- $5^\circ$, respectively, for the hemisphere illuminated by the GRB at zenith point $90^\circ$ N.  These profiles are then interpolated and the zenith frame latitudes are transformed into standard Earth coordinates (longitude and latitude) according to the impact latitude chosen for the event.  Finally, a zonally averaged profile (as a function of altitude and latitude) is produced by averaging over all longitudes for each $10^\circ$ latitude band.  This profile is then input to the atmospheric model as a production source of $\mathrm{NO_y}$ and $\mathrm{HO_x}$, as described above.

\subsection{Simulation Runs}
\label{model-sims}
Our simulation runs are performed as follows.  Initial conditions are obtained from long-term runs (roughly 40 years) intended to bring the model to equilibrium.  These runs end at the various times of year at which we input the GRB.  Constituent values from these equilibrium runs are read in by the $1~\mathrm{s}$ time step version of the model which runs for 7 days, beginning and ending at noon.  Ionization due to the gamma-rays (as calculated off-line, see Sect.~\ref{model-grbinput}) is input as a step function (top hat) with duration $10~\mathrm{s}$.  Most runs input the burst at noon on day 4.  Bursts at other times of day occur after this.  We use day 4 as the input point in order to allow the short-term model to come to equilibrium.  This is necessary as the two versions have different time scales for chemistry ($1~\mathrm{s}$ versus daily averages) and differences in transport.  The run then continues for three more days to allow for good time resolution of the chemistry directly following the burst.  Running for much longer than 7 days would introduce errors due to the limited transport in the short term model.  Hence, at the end of 7 days we then feed constituent values to the long term (1 day time step) version of the model which runs for 20 years, by which time the atmosphere has recovered from the GRB perturbation.  Runs are also performed using the same steps without input of GRB ionization, in order to make comparisons between perturbed and unperturbed runs (e$.$g$.$ $\mathrm{O_3}$ percent change).

\subsection{Biological Effects}
\label{model-bio}
In order to quantify some of the effects on life due to a GRB impact, we have computed the direct effect of the increased solar UVB (due to depletion of ozone) by determining the irradiance in this waveband at the surface of the Earth and convolving it with a biological weighting function which describes the effectiveness of particular wavelengths in damaging DNA molecules.

\begin{figure}
\plotone{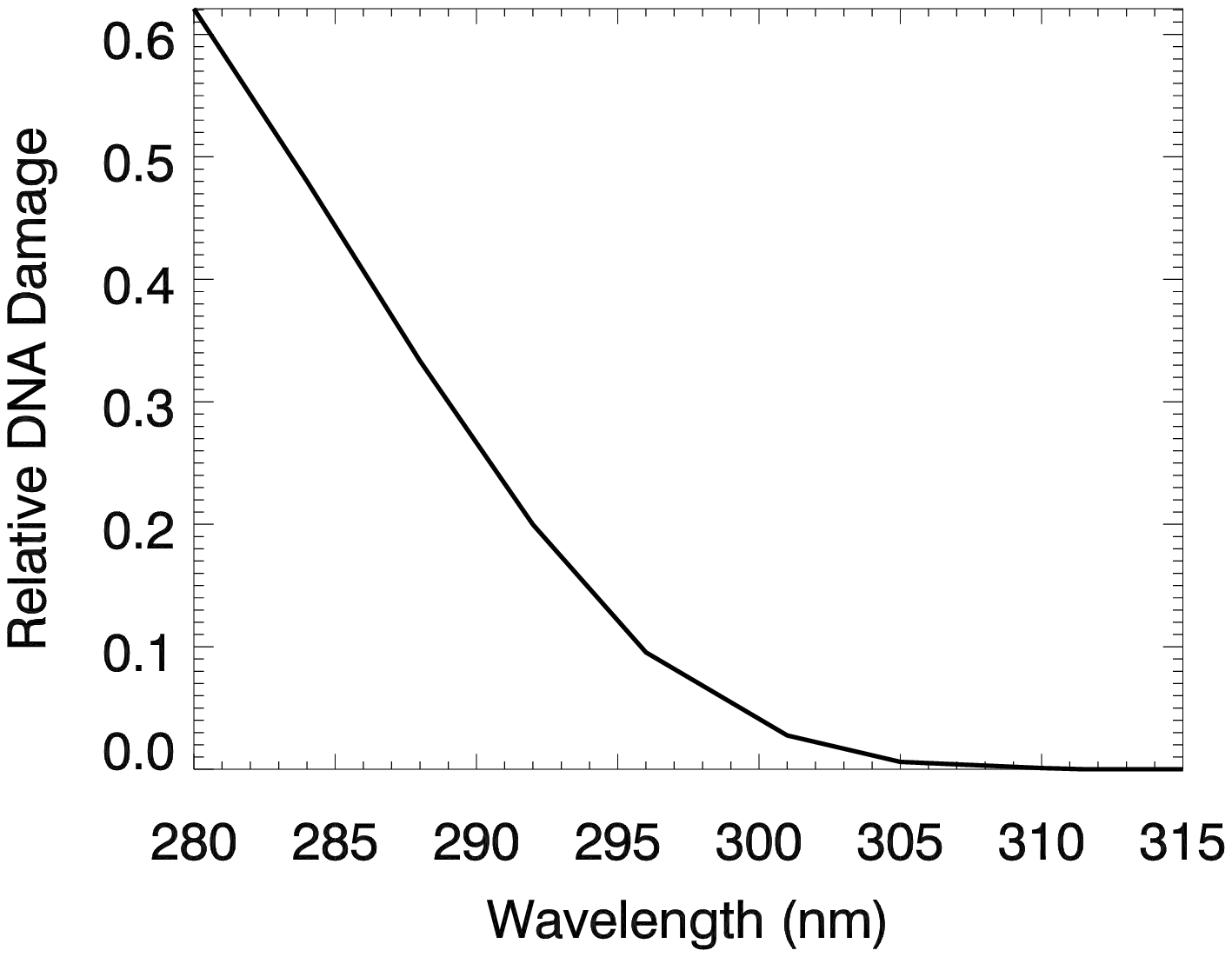}
\caption{Biological weighting function of \citet{set74} and \citet{smith80}, quantifying damage to DNA by UVB.
\label{fig:DNAweight}}
\end{figure}

In order to determine the solar UVB reaching the surface, we take the irradiance at the top of the atmosphere (obtained from data taken by the SUSIM ATLAS instrument, available: $\mathrm{http://wwwsolar.nrl.navy.mil/susim.html}$) and determine the irradiance at the surface (for a given wavelength band) using the Beer-Lambert Law, 
\begin{equation}
I = I_0 e^{-\sigma(\lambda) N / \cos\theta}
\end{equation}
where $\sigma(\lambda)$ is the absorption cross section of ozone as a function of wavelength, $\lambda$, (e$.$g$.$ \citet{yos88}), $N$ is the vertical column density of ozone (taken from our modeling results) and $\theta$ is the solar zenith angle (calculated for a particular latitude, day of year and time of day) \citep{mad93}.  Only ozone absorption effects are included in this calculation since the effect of scattering at these wavelengths is comparatively small (see Sect.~\ref{uncert-UV}).  The surface irradiance is then multiplied by the value of the biological weighting function (\citet{set74,smith80}; see Fig.~\ref{fig:DNAweight}) for that wavelength band and this combined DNA damage value is then integrated over the entire wavelength range.  This value is then scaled by $\cos\theta$ to account for insolation.  The above calculation is done for each hour and then averaged over 24 to get a daily average DNA damage value for a given latitude and time point.

\section{Results} 
\begin{figure}
\plotone{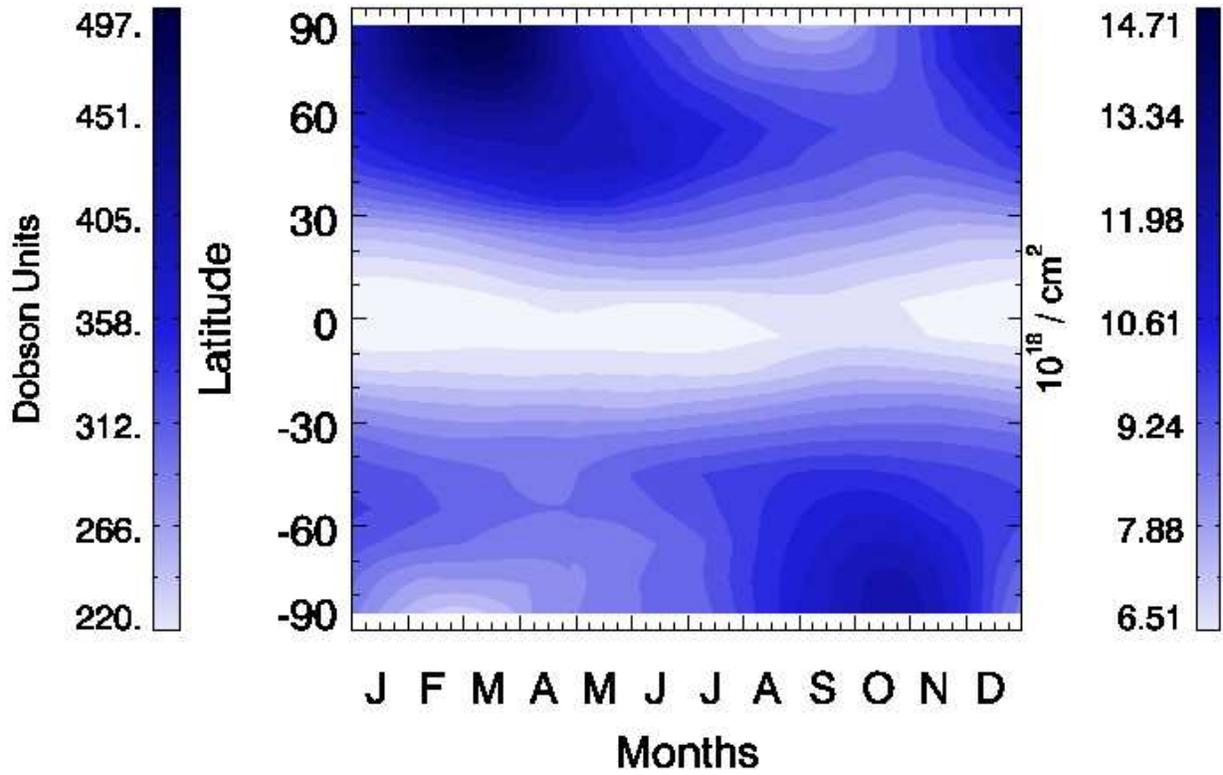}
\caption{Annual variation of the column density of $\mathrm{O_3}$ in an unperturbed run, with scales for both Dobson units (left) and $10^{18}~\mathrm{cm^{-2}}$ (right).  The horizontal axis labels give month initials starting in January.  (See color figure online.)
\label{fig:O3_coldens-NOgrb}}
\end{figure}

The primary atmospheric results of our simulations are increases in $\mathrm{NO_y}$ and decreases in $\mathrm{O_3}$.  Ozone column densities computed in the model are then used to calculate the UVB flux at the Earth's surface and the resulting DNA damage.  For purposes of reference and validation of our model's ability to correctly simulate normal ozone variations we show in Fig.~\ref{fig:O3_coldens-NOgrb} a plot of $\mathrm{O_3}$ column density for the unperturbed atmosphere as a function of latitude and time, over one year, starting in January.  Units shown are both Dobson units and $10^{18}~\mathrm{cm^{-2}}$.  A Dobson unit describes the thickness of a column of ozone at STP and is defined as $1~\mathrm{DU} = 0.01~\mathrm{mm}$ thickness (or, $1~\mathrm{DU} = 2.69\times10^{18}~\mathrm{cm^{-2}}$).  CFC's and other anthropogenic constituents are not included in our runs (see Sect.~\ref{app:model}), so the values and distribution of $\mathrm{O_3}$ in this plot differ somewhat from those in the present day atmosphere.  Anthropogenic constituents are excluded because we are primarily interested in applying our results to pre-industrial time periods.

Results for an initial set of parameters were reported in \citet{thom05}.  We repeat those results here, in the context of a wider exploration of parameters.  
First, we discuss the effects of varying the time of year and the latitude at which the burst occurs for a given fluence (Sect.~\ref{results-100kJ}).  Then, we discuss the effects of the burst occurring at different times of day and with various durations, for a given fluence (Sect.~\ref{results-time}).  Next, we discuss the effects of varying the burst fluence (Sect.~\ref{results-10kJ}).  We also present results quantifying the impact of reduced $\mathrm{O_3}$ levels on life (Sect.~\ref{results-bio}).  Finally, we discuss the reduction of sunlight due to increased $\mathrm{NO_2}$ levels and the deposition of nitrates due to rainout of $\mathrm{HNO_3}$ (Sect.~\ref{results-opacity}).  Note that most figures presented here are available in color in the online version of the paper.

\subsection{The Importance of Sunlight}
\label{results-sun}
As will be seen throughout the discussion that follows, the presence or absence of sunlight at a given location and time has important implications for our atmospheric results.  This is due to photolytic reactions involving $\mathrm{O_3}$ and constituents of $\mathrm{NO_y}$, most importantly NO and $\mathrm{NO_2}$.  

The primary creation mechanism for $\mathrm{O_3}$ is by dissociation of $\mathrm{O_2}$ by solar UV at wavelengths below 242 nm.  Higher wavelengths dissociate $\mathrm{O_3}$, which is why ozone provides an effective shield at these wavelengths.  A consequence of the photo-production of $\mathrm{O_3}$ is that ozone is primarily created in the tropics, where sunlight is most constant and intense.  Of course, transport in the upper atmosphere is primarily from the tropics to the poles, and so ozone is moved to the poles after being produced near the equator.  Longitudinal transport is quite efficient and only takes a few weeks to circle the globe.

Photolysis is also important for $\mathrm{NO_y}$ constituents and their effects on ozone.  NO and $\mathrm{NO_2}$, the primary ozone depleting components of $\mathrm{NO_y}$, are dissociated by sunlight at wavelengths below 191 and 400 nm, respectively. In addition, other $\mathrm{NO_y}$ constituents are photolyzed to produce NO and $\mathrm{NO_2}$.  (See Table~\ref{tbl:rxns3} for photolysis reactions and Table~\ref{tbl:bands1} for solar fluxes included in the model.)  Photolysis of $\mathrm{NO}$ has two important implications for both $\mathrm{NO_y}$ and $\mathrm{O_3}$.  This reaction yields atomic N which can reduce NO through the reaction $\mathrm{N + NO \rightarrow N_2 + O}$, thereby decreasing concentrations of $\mathrm{NO_y}$.  In addition, atomic O is produced which can actually increase $\mathrm{O_3}$ through reaction with  $\mathrm{O_2}$.

The diurnal cycle of NO and $\mathrm{NO_2}$ exhibits an interchange between these compounds.  NO concentrations decrease as the sun sets due to the removal of its photolysis source and its rapid oxidation to $\mathrm{NO_2}$.  During the day, NO is created partly through photolysis of $\mathrm{NO_2}$ thereby reducing the amount of $\mathrm{NO_2}$ present.  $\mathrm{NO_2}$ is more strongly affected by photolysis than NO, since its upper wavelength limit (400 nm) is higher than that for NO (191 nm) and there are more photons present in the solar flux below 400 nm than 191 nm (Table~\ref{tbl:bands1}).

\begin{figure}
\plotone{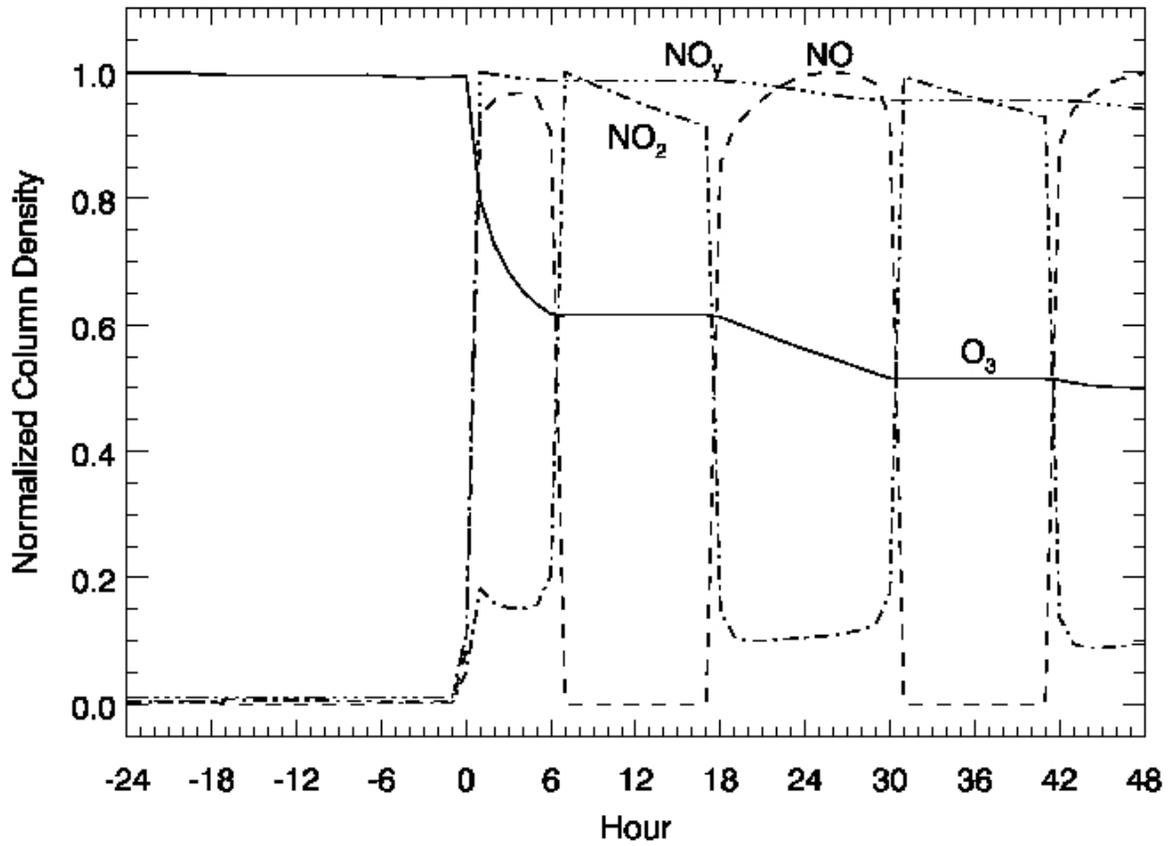}
\caption{
Column density of NO, $\mathrm{NO_2}$, $\mathrm{NO_y}$, and $\mathrm{O_3}$, normalized to their maximum values over this time period.  These are hourly results from the short-term version of the model for a $100~\mathrm{kJ/m^2}$ burst over the equator in late March at noon.  Sudden changes (particularly in NO and $\mathrm{NO_2}$) occur at sunrise/sunset.  (Burst occurs at time 0, which is local noon.)
\label{fig:diurnal}}
\end{figure}

These interchanges can be seen in Fig.~\ref{fig:diurnal} where we plot column densities of several constituents at $-5^{\circ}$ latitude, normalized to their maximum values over this time period.  This data is for a $100~\mathrm{kJ/m^2}$ burst over the equator in late March at noon.  Time 0 indicates the burst input.  It is apparent from this plot that the burst causes an immediate increase in the concentration of $\mathrm{NO_y}$ (primarily through NO).  After the burst, normal diurnal processes dominate the relative amounts of these compounds at any given time, though at much higher levels than normal, and with the addition of the long term depletion of $\mathrm{O_3}$.

The effects of sunlight on NO, $\mathrm{NO_2}$ and $\mathrm{O_3}$ are important for understanding many features of our results and we will refer to them throughout the following discussion.

\subsection{Varying Time of Year and Latitude}
\label{results-100kJ}
\begin{figure}
\plotone{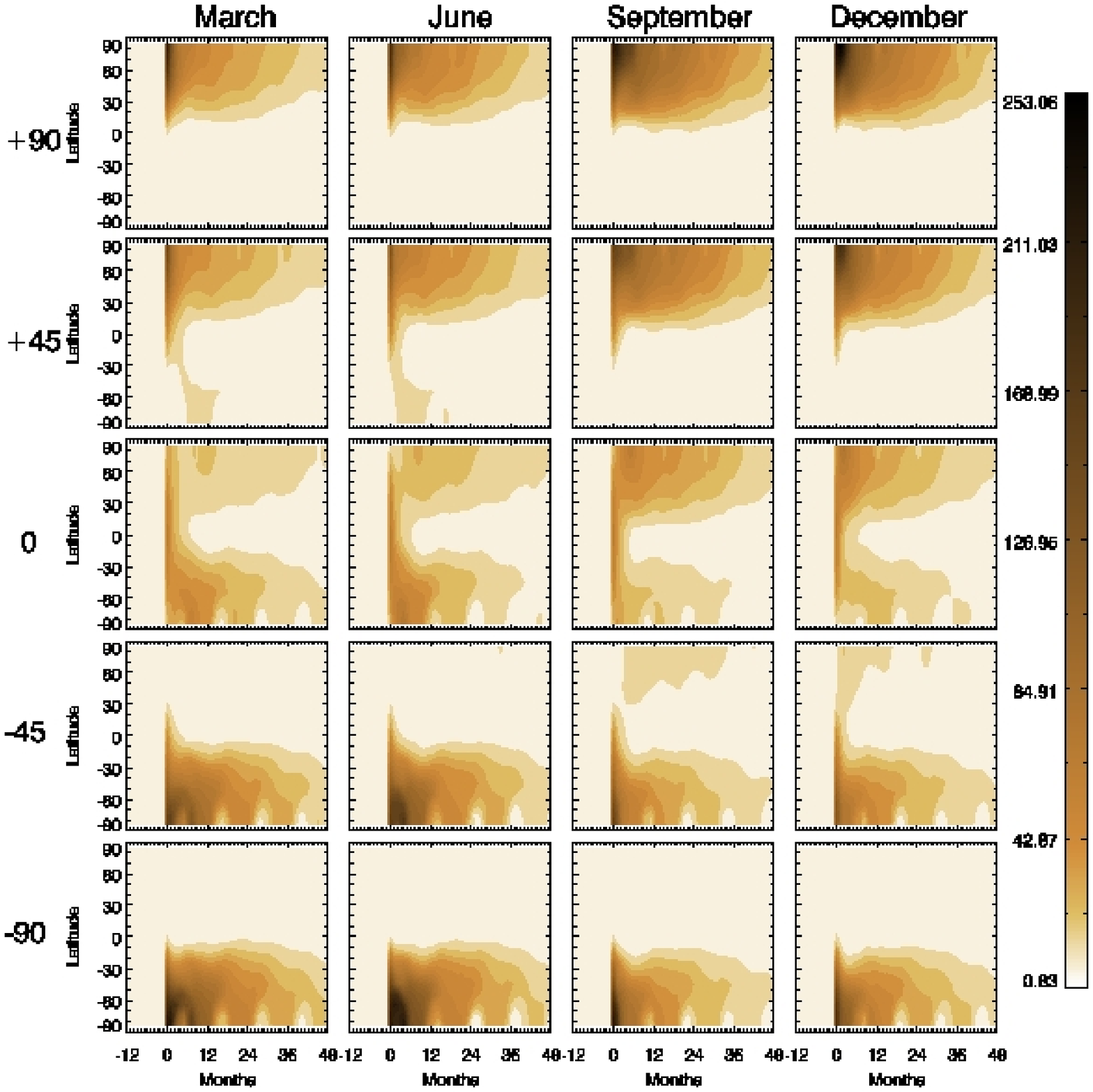}
\caption{Column density of $\mathrm{NO_y}$ in units of $10^{16}~\mathrm{cm^{-2}}$, for $100~\mathrm{kJ/m^2}$ bursts over latitudes $+90^\circ$, $+45^\circ$, the equator, $-45^\circ$, and $-90^\circ$, at the equinoxes and solstices. (Bursts occur at month 0. See color figure online.)
\label{fig:NOy_coldens}}
\end{figure}

\begin{figure}
\plotone{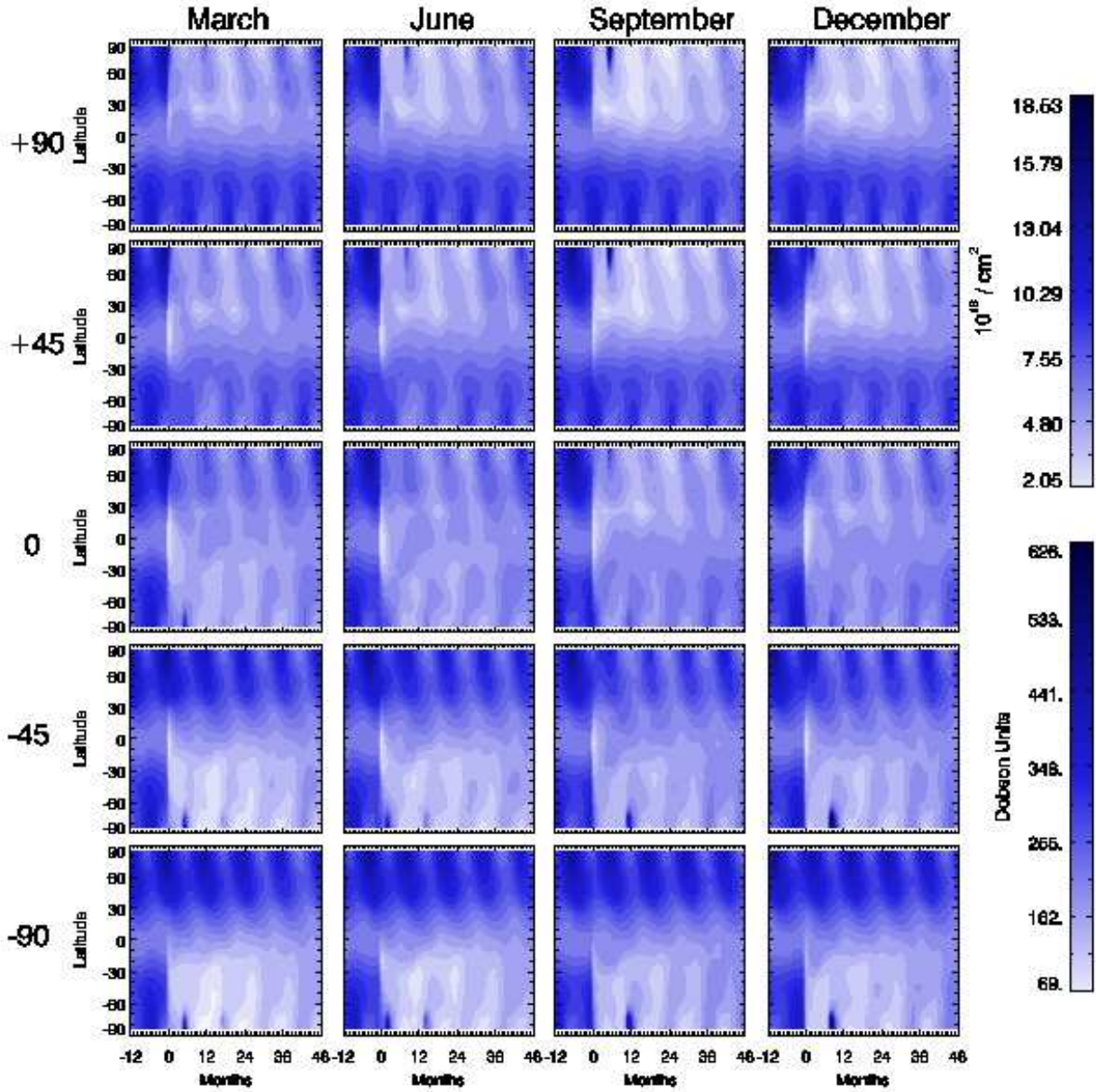}
\caption{Column density of $\mathrm{O_3}$ with scales for both Dobson units (left) and $10^{18}~\mathrm{cm^{-2}}$ (right), for $100~\mathrm{kJ/m^2}$ bursts over latitudes $+90^\circ$, $+45^\circ$, the equator, $-45^\circ$, and $-90^\circ$, at the equinoxes and solstices. (Bursts occur at month 0. See color figure online.)  
\label{fig:O3_coldens}}
\end{figure}

\begin{figure}
\plotone{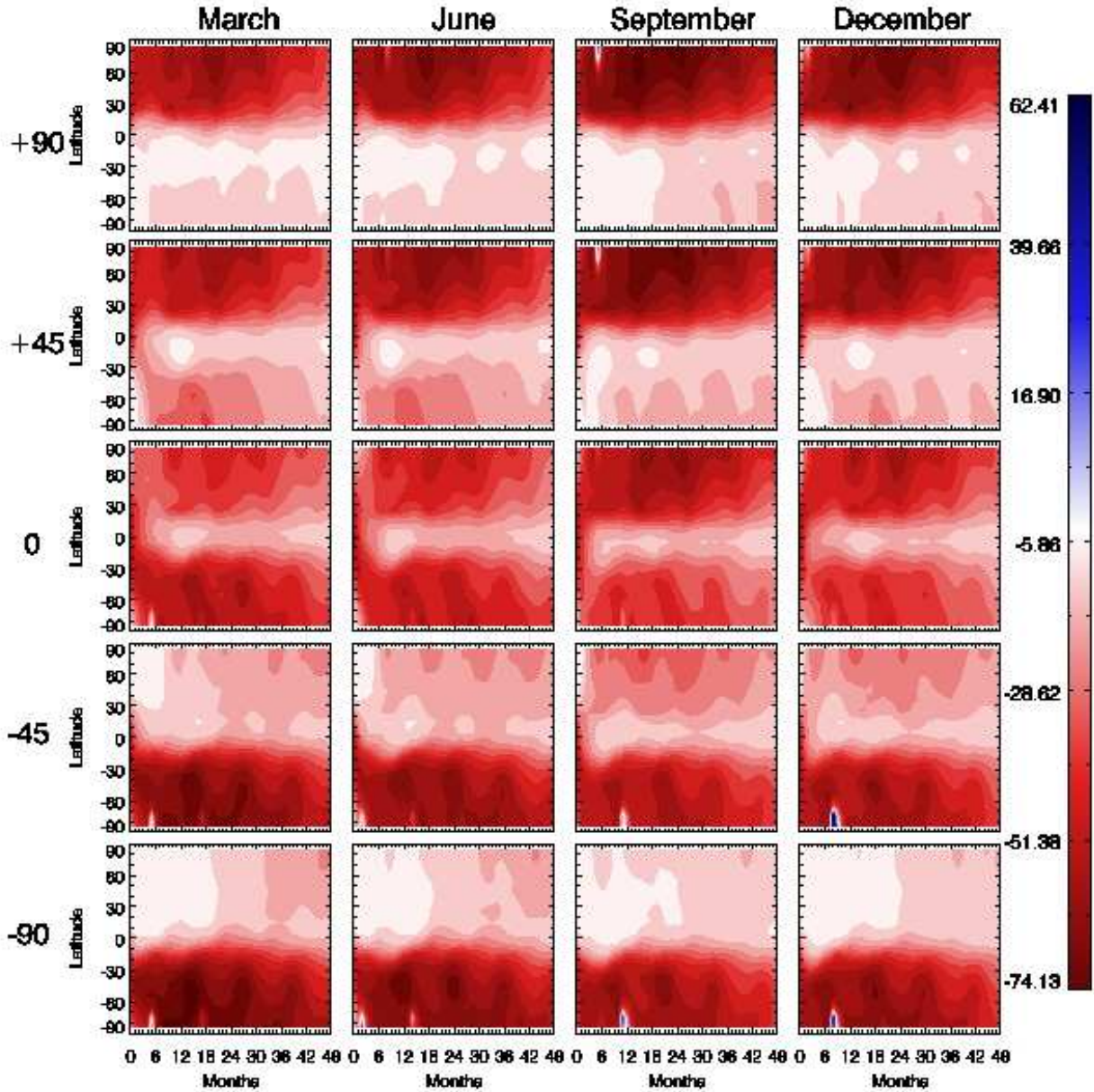}
\caption{Pointwise percent change in column density of $\mathrm{O_3}$ (comparing runs with and without burst), for $100~\mathrm{kJ/m^2}$ bursts over latitudes $+90^\circ$, $+45^\circ$, the equator, $-45^\circ$, and $-90^\circ$, at the equinoxes and solstices. (Bursts occur at month 0.)  Note that in this plot white indicates values equal to or greater than 0.0.  The ``wedges'' of white that appear in the polar regions in some frames are the only areas of positive percent change.  See color figure online.
\label{fig:O3_perchg}}
\end{figure}

\begin{figure}
\plotone{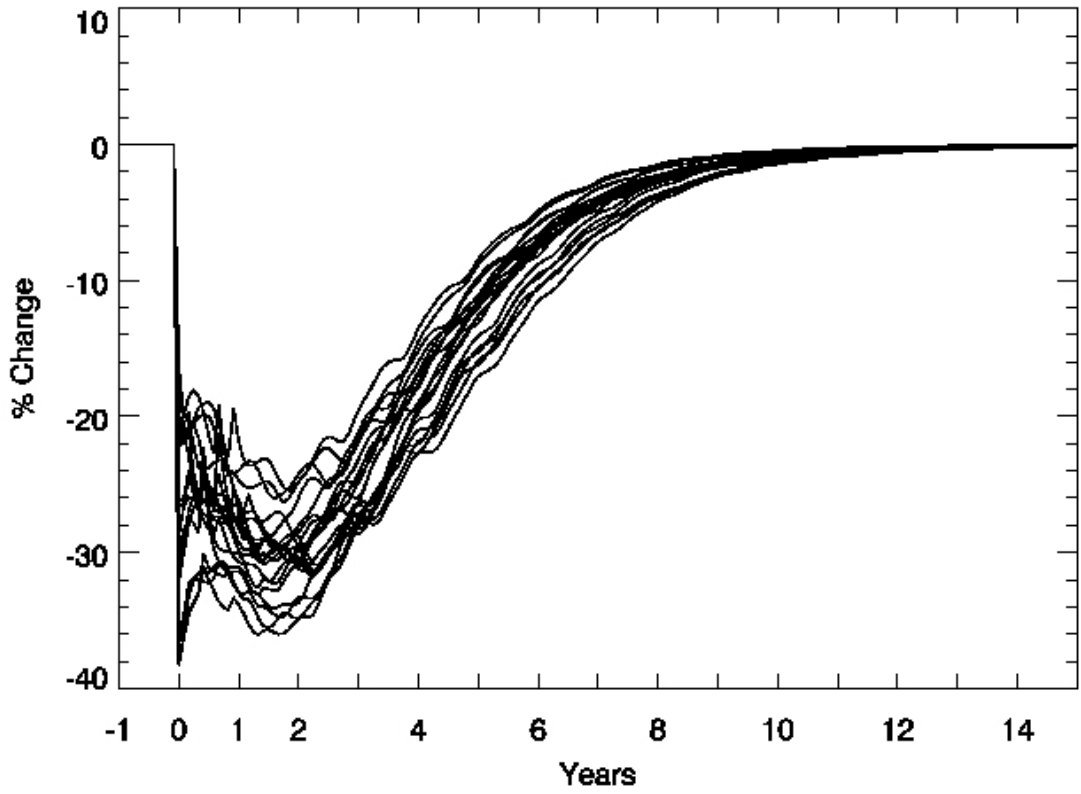} 
\caption{Percent change in globally averaged column density of $\mathrm{O_3}$, for $100~\mathrm{kJ/m^2}$ bursts over latitudes $+90^\circ$, $+45^\circ$, the equator, $-45^\circ$, and $-90^\circ$, at the equinoxes and solstices. (Bursts occur at month 0.)
\label{fig:O3_perchg-glob}}
\end{figure}

We will now discuss results similar to those presented in \citet{thom05} for bursts occurring at the equinoxes and solstices, over five different latitudes.
Figures~\ref{fig:NOy_coldens}, \ref{fig:O3_coldens}, \ref{fig:O3_perchg}, and \ref{fig:O3_perchg-glob} show results for bursts of fluence $\mathrm{100~kJ/m^2}$ input in late March (repeating the results from \citet{thom05}), June, September and December, at latitudes $+90^\circ$, $+45^\circ$, $0^\circ$ (equator), $-45^\circ$, and $-90^\circ$.  For all these cases the burst is input at noon.  In each plot, the burst is input at time 0. 

Figure~\ref{fig:NOy_coldens} shows the vertical column density of $\mathrm{NO_y}$, in units of $10^{16} ~\mathrm{cm^{-2}}$, at each latitude point in the model over time for each combination of event time of year and latitude. 
Similarly, Fig.~\ref{fig:O3_coldens} shows the vertical column density of $\mathrm{O_3}$.  Included are scales in both Dobson units and $10^{18}~\mathrm{cm^{-2}}$.  
%
Figure~\ref{fig:O3_perchg} shows the percent difference at a given location and time (between a run with gamma-ray input and one without) in vertical column density of $\mathrm{O_3}$.  (Note that in this plot white indicates values equal to or greater than 0.0.  The ``wedges'' of white that appear in the polar regions in some frames are the only areas of $\mathrm{O_3}$ \textit{increase}.  This will be discussed further below.)
The full long-term duration of the effects can be seen in Fig.~\ref{fig:O3_perchg-glob}, which shows the global average percent change in ozone as a function of time for 15 years after the burst (with one year of 0 change before the burst).  
We summarize the maximum depletion of ozone (globally averaged and at a given latitude) in Tables~\ref{tbl:1} and \ref{tbl:2}.

\begin{figure}
\plotone{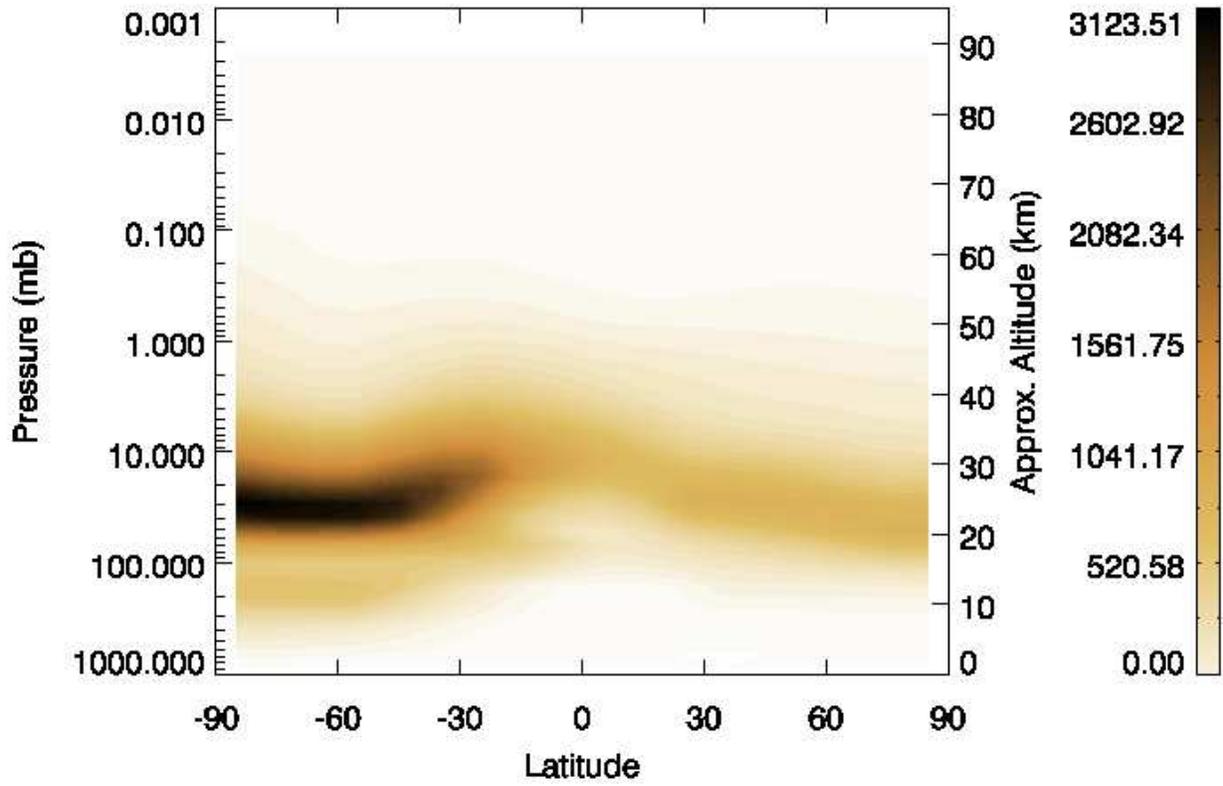}
\caption{Pointwise percent change in number density of $\mathrm{NO_y}$ (comparing runs with and without burst), for $100~\mathrm{kJ/m^2}$, March, equator case, 12 months after burst. (See color figure online.)
\label{fig:NOy_perchg_altlat}}
\end{figure}

\begin{figure}
\plotone{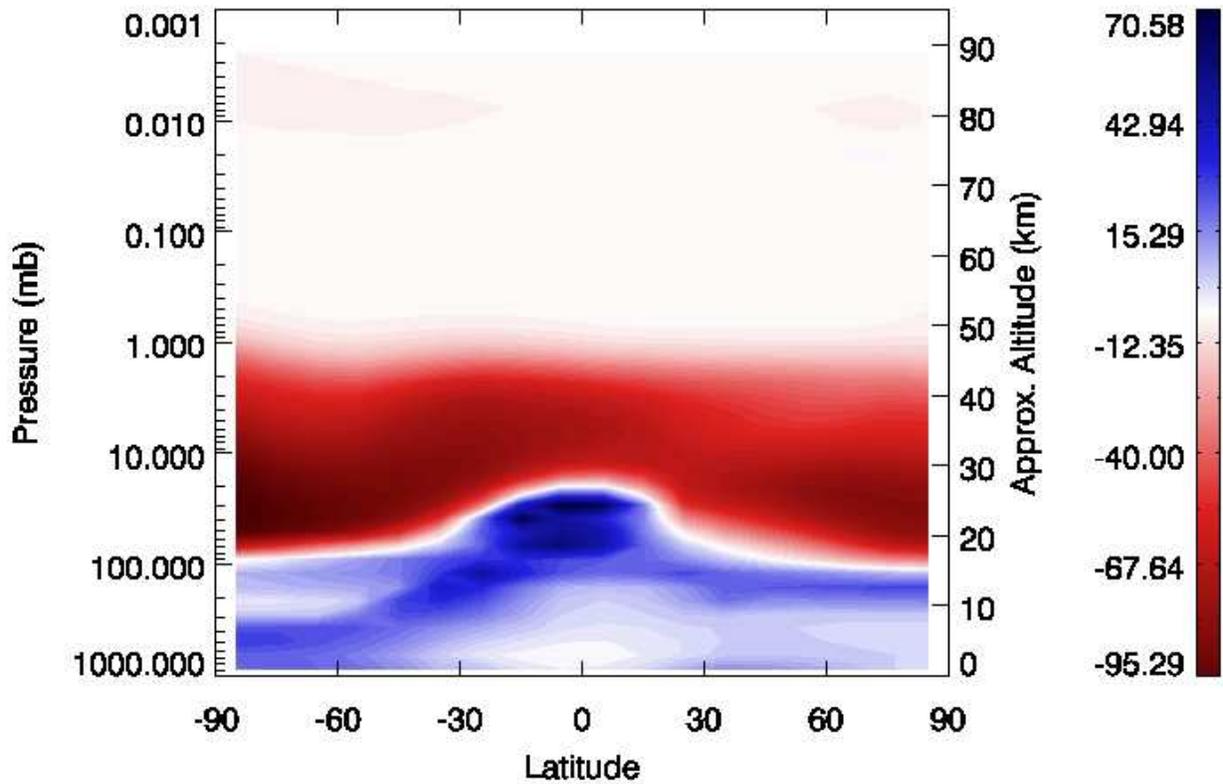}
\caption{Pointwise percent change in number density of $\mathrm{O_3}$ (comparing runs with and without burst), for $100~\mathrm{kJ/m^2}$, March, equator case, 12 months after burst.  Black in this figure represents both positive and negative percent changes.  White is set to 0.0. The white ``boundary'' starting at about 15 km in the polar regions and moving upward toward the equator separates areas of negative percent change at higher altitudes and positive percent change at lower altitudes.  See color figure online.
\label{fig:O3_perchg_altlat}}
\end{figure}

To give a sense of the distribution of atmospheric effects with altitude we also include Fig.~\ref{fig:NOy_perchg_altlat} and \ref{fig:O3_perchg_altlat}, which show pointwise percent difference in $\mathrm{NO_y}$ and $\mathrm{O_3}$ between a run with a burst and one without, at each altitude and latitude at one year after the burst occurs.  These plots are for a burst in late March over the equator.  

An interesting feature apparent in Fig.~\ref{fig:O3_perchg_altlat} is that while ozone levels show a decrease at altitudes above about 20 km, they are actually increased at or below this altitude, particularly around the equator.  (The white ``boundary'' starting at about 15 km in the polar regions and moving upward toward the equator separates areas of $\mathrm{O_3}$ decrease at higher altitudes and $\mathrm{O_3}$ increase at lower altitudes.)  This production at lower altitudes is due to the following.  Normally, ozone is concentrated between 30 and 40 km altitude and is sparse at lower altitudes.  However, the burst depletes ozone primarily in this altitude range, which allows solar UV (normally absorbed by the ozone) to penetrate to lower than normal altitudes.  This UV then creates ozone (by photolysis of $\mathrm{O_2}$) at altitudes where normally there is little or none.  (\citet{jack85} discuss observation of this effect in the case of a solar proton event.)  This effect is particularly strong around the equator, since the intensity of solar UV is greatest there.  Therefore, since this plot shows a comparison between a run without a burst (where little ozone exists at lower altitudes) and one with a burst (where UV penetrates more deeply and creates lower altitude ozone), we see an increase at these lower altitudes.  This effect actually somewhat mitigates the total column depletion.

Several important features are apparent in Figs.~\ref{fig:NOy_coldens}--\ref{fig:O3_perchg-glob}.  First, immediate atmospheric effects are seen, i$.$e$.$ production of $\mathrm{NO_y}$ and depletion of $\mathrm{O_3}$, beginning around the latitude over which the event occurs and spreading quickly to other latitudes.  Note that for polar bursts, even in the long term, effects are isolated to the respective hemisphere.  Similarly, even for bursts at $\pm45^\circ$ the opposite hemisphere experiences much less intense effects.  This is due to the fact that transport in the atmosphere is primarily poleward from the equator and transport across the equatorial line is minimal.  Therefore, $\mathrm{NO_y}$ produced in one hemisphere is for the most part not transported to the opposite hemisphere.  

Isolation of effects to a hemisphere means that polar and $\pm45^\circ$ bursts result in smaller globally averaged ozone depletions as compared to bursts over the equator in most cases (see Table~\ref{tbl:1}).  This is true even though maximum localized depletions are typically larger for polar bursts (Table~\ref{tbl:2}).  
Local intensity is understandable in these cases, since most of the $\mathrm{NO_y}$ is produced directly in the polar regions and subsequently is disbursed over a smaller area.
This high concentration contained within the polar region allows a greater effect on ozone.

For all cases, ozone depletion is long-lived (Fig.~\ref{fig:O3_perchg-glob}).  Globally averaged depletions of at least 10\% are present up to 5 -- 7 years and full recovery is not achieved until 10 -- 12 years after the burst.  Recovery occurs at the equator first.  This is normal in the terrestrial atmosphere, as ozone is primarily produced in tropical latitudes and transported poleward.  Also, as discussed above, $\mathrm{NO_y}$ produced at the equator is moved toward the poles fairly rapidly, so concentrations are significantly reduced at the equator (reducing the depletion of ozone there) within a year or so, even for equatorial bursts.

It is interesting to note that for all the cases discussed so far recovery of ozone takes approximately a decade.  Why this time scale and not some other?  The primary controlling factor here is the lifetime of the ozone depleting $\mathrm{NO_y}$ compounds which are generated by the burst.  These compounds are removed from the atmosphere mainly through transport down to the troposphere where they are rained out.  \citet{flem99} and \citet{flem01} compare results of the GSFC model with measurements of the ``age of air'' as determined from measurements of long-lived gases such as $\mathrm{SF_6}$ and $\mathrm{CO_2}$.  This ``age of air'' at the tropopause is on the order of 5 years.  Only after concentrations of  $\mathrm{NO_y}$ in the stratosphere are significantly reduced by this transport process can  $\mathrm{O_3}$ levels recover, a process which is also dominated by transport and hence occurs on a similar timescale to the removal of $\mathrm{NO_y}$.  Therefore, the combination of removal of $\mathrm{NO_y}$ and replacement of $\mathrm{O_3}$ results in a total recovery time of approximately a decade.

Greater depletion of ozone in the polar regions is a general feature of these results and is observed in the present day atmosphere as well.  
%
%
%
Figure~\ref{fig:O3_perchg} gives a somewhat exaggerated sense of the effect of depletion at the poles because ozone is initially high there 
(see Figs.~\ref{fig:O3_coldens-NOgrb} \ref{fig:O3_coldens}).  
Larger ozone depletions at the poles are due to the preferential transport of $\mathrm{NO_y}$ poleward, combined with the long lifetime of the enhanced $\mathrm{NO_y}$ in the polar stratosphere.  This long lifetime is primarily due to lower photolysis rates at high latitudes (from low solar incidence angle and dark polar winters; see Sect.~\ref{results-sun}).  

The enhanced $\mathrm{NO_y}$ in the polar regions, including $\mathrm{HNO_3}$, leads to an enhancement of nitric acid trihydrate (NAT) polar stratospheric clouds (PSCs).  These NAT PSCs facilitate heterogeneous reactions that result in greater ozone depletion by halogen (chlorine and bromine) constituents.  
The contribution of PSCs is especially strong in the south polar region where a stronger polar vortex with colder stratospheric temperatures is in place during the winter.  This leads to the south polar ozone ``hole'' which appears annually in the present day atmosphere.

An effect of PSCs is seen in Fig.~\ref{fig:NOy_coldens} where column density of $\mathrm{NO_y}$ is reduced during polar winter, especially at the south pole. (The feature is present in the north, but is less intense.)  This is due to the fact that during polar night most of the $\mathrm{NO_y}$ is in the form of $\mathrm{HNO_3}$ and, particularly at the south pole, this $\mathrm{HNO_3}$ is largely in solid form in PSCs.  This form of $\mathrm{HNO_3}$ is not included in the summation of total $\mathrm{NO_y}$ and so the $\mathrm{NO_y}$ column density shown here appears reduced during these times.  Note, however, that it returns to high values at the end of polar night, when solid $\mathrm{HNO_3}$ is converted back to $\mathrm{NO_2}$.  Over the long term, the column density of $\mathrm{NO_y}$ which ``reappears'' after polar night is less than it was before.  This is because some PSC particles fall down to the troposphere where $\mathrm{HNO_3}$ is rained out.  This is the primary mechanism whereby $\mathrm{NO_y}$ is permanently removed from the atmosphere.

While the particular atmospheric conditions which lead to the north-south asymmetry seen here are peculiar to the present-day configuration of continents, the effect of varying burst latitude and time of year are much larger.  For example, the asymmetry is reversed for an equatorial burst in September, and the strength of the asymmetry is less for equatorial bursts in June and December.

\begin{figure}
\plotone{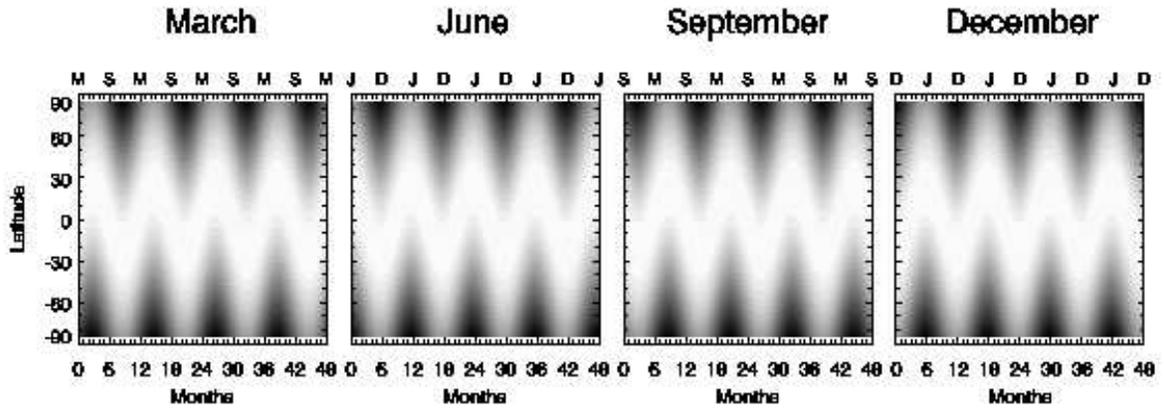}
\caption{Intensity of sunlight at noon (in arbitrary units), beginning at the four times of year at which we have input bursts.  On the upper axes, M = March, J = June, S = September, D = December. \label{fig:sunup}}
\end{figure}

The relation of the time of year of the burst to the season at the polar regions is an essential factor in the variation of several features between the different burst cases.  This includes differences such as the shift in north-south asymmetry as well as variation of maximum ozone depletion (both globally averaged and localized) for bursts over a given latitude which occur at different times of year.  Other features are affected by this timing as well and will be discussed later in this section. 
The key variable is how much time elapses between when the burst occurs and when polar night ends. 
Figure~\ref{fig:sunup} shows (in arbitrary units) the intensity of sunlight (at noon) at each latitude over 4 years,  starting in late March, June, September and December, similar to Figs.~\ref{fig:NOy_coldens}--\ref{fig:O3_perchg}.  This figure may allow the reader to more easily relate the solar intensity at a given time of year and latitude and the effects to be discussed below.

Relative timing of the burst and polar spring is important due to the fact that photolysis is an important factor in the chemistry of $\mathrm{NO_y}$ and ozone and hence the presence (and incidence angle) of sunlight is critical (see Sect.~\ref{results-sun}).  $\mathrm{NO_y}$ constituents, particularly $\mathrm{NO}$ and $\mathrm{NO_2}$ (the most important for $\mathrm{O_3}$ depletion), are strongly photolyzed.  Therefore whether the Sun is up in the polar regions, where $\mathrm{NO_y}$ tends to concentrate, is important for how much of an effect on ozone these compounds have.  Relative timing of the production of $\mathrm{NO_y}$ by the burst and the appearance of the Sun after polar winter is the primary cause of differences in distribution and intensity of ozone depletion for bursts at different times of year.  Important here is the fact that it takes some time for $\mathrm{NO_y}$ to be transported to the polar regions when the burst is equatorial.  Since the removal time for $\mathrm{NO_y}$ is on the order of several years, these compounds build up in the polar regions.

For instance, for an equatorial burst in March, high southern latitudes are in darkness for 5--6 months after the burst (depending on latitude), whereas at high northern latitudes the 
Sun is up for most of the day during that time period and the next polar spring does not begin for some 11--12 months.  This means that $\mathrm{NO_y}$ produced at the equator and transported to the northern polar regions may be reduced through photolysis and subsequent reactions while that transported to the southern polar regions is not.  Hence, the concentration of $\mathrm{NO_y}$ is higher for longer in the high southern latitudes (see Fig.~\ref{fig:NOy_coldens}).  Note that the opposite scenario is true for an equatorial burst in September, when the relative timing of the burst and spring at the poles is reversed.  This may also be seen in the ozone depletion resulting from the concentration of $\mathrm{NO_y}$ (Figs.~\ref{fig:O3_coldens} and \ref{fig:O3_perchg}).

For bursts which occur in June and December, $\mathrm{NO_y}$ is somewhat more evenly distributed with latitude and ozone depletion effects are reduced, as compared to bursts in March and September.  For these cases, the time between when the burst occurs and the soonest polar spring is smaller for both poles.  If polar spring occurs too soon after the burst, significant amounts of $\mathrm{NO_y}$ will not be transported into the region before night returns and so the reduction of concentration by photolysis will be lessened.  This is apparent for an equatorial burst in June at high southern latitudes and in December at high northern latitudes.

For bursts over the poles and $\pm45^\circ$ latitude there are likewise variations in ozone depletion with burst timing, though primarily in intensity rather than distribution, since  for these bursts most of the $\mathrm{NO_y}$ produced is contained within a hemisphere, as discussed earlier.  Consider a north polar burst.  The smallest depletion of ozone occurs for a burst in March.  This is due to the fact that the Sun is up at most high latitudes at this time and is for several months after the burst.  For a June burst the depletion becomes somewhat greater, since north polar night begins sooner after the burst.  A December burst has yet higher depletion; here the burst occurs in the middle of north polar night.  The greatest depletion occurs for a burst in September.  In this case, the burst occurs just as north polar night is beginning.  Note that the relative levels of depletion are reversed between March and September and June and December for a south polar burst, simply due again to the relation of the time of year of the burst and the season at high southern latitudes.  

Similar relationships are present for the bursts over $\pm45^\circ$ latitude.  Here there is the additional feature of how much depletion occurs in the opposite hemisphere.  For instance, a March burst at $-45^\circ$ exhibits very little ozone depletion in the northern hemisphere, while a September burst results in a fair amount of depletion in the northern hemisphere, but less in the southern, as compared to the March case.  The same phenomena described above apply here as well.  In these cases, of course, most $\mathrm{NO_y}$ is produced and remains in the hemisphere over which the burst occurs, as in the polar cases.  However, since the burst occurs at mid latitudes, some $\mathrm{NO_y}$ is produced at the equator or even in the opposite hemisphere.

A feature which is obvious in Fig.~\ref{fig:O3_perchg} is the presence of brief ``spikes'' of ozone \emph{production}.  These spikes appear with varying intensity for most burst cases.  Some occur as lower depletion than their surroundings, while some exhibit an actual increase in ozone as compared to conditions without a burst.  All occur close to the poles and are short lived (1--2 months).  The presence and intensity of these spikes are related to the relative time of year of the burst and the season at the polar regions, similar to the effects discussed above.

Ozone production at the start of polar spring in the presence of high concentrations of $\mathrm{NO_y}$ can be understood as follows.  As discussed above, $\mathrm{NO_y}$ exists primarily as $\mathrm{HNO_3}$ during polar night.  When the Sun suddenly appears, $\mathrm{NO_2}$ is produced by photolysis of $\mathrm{HNO_3}$.  $\mathrm{NO_2}$ is then photolyzed to NO and O and NO is further photolyzed producing additional O. Direct production of $\mathrm{O_3}$ can then occur when O reacts with $\mathrm{O_2}$.  In addition, 
decreasing amounts of PSCs (as the temperature rises) and increasing amounts of $\mathrm{NO_2}$ reduces the ozone depleting action of chlorine and bromine compounds. ($\mathrm{NO_2}$ reacts with ClO and BrO, sequestering Cl and Br in $\mathrm{ClONO_2}$ and $\mathrm{BrONO_2}$ which do not directly interact with ozone.)


\subsection{Varying GRB Duration and Time of Day}
\label{results-time}
We will now discuss the differences in results due to varying the time of day at which the burst occurs.  All the results presented so far are for a burst input at noon.  The short duration of the burst combined with the normal variation of $\mathrm{O_3}$ and $\mathrm{NO_y}$ constituents (especially NO and $\mathrm{NO_2}$) over the day-night cycle make it reasonable to expect that bursts which occur at different times of day, especially during daylight versus dark, will lead to differing results.  

In Table~\ref{tbl:3} we present the maximum ozone depletions, globally averaged and localized, for bursts occurring every two hours during the day and night, starting at noon, in late March, over the equator.  As can be seen, there is a difference of 3 to 4 percent between daytime and nighttime bursts.  However, those which all occur during daylight hours or all occur during dark hours are nearly identical to each other.  We have also investigated a burst at midnight over the equator in late June, which is the maximally different case from March, in terms of season.  In this case, the maximum globally averaged and localized ozone depletion is -40\% and -59\%, respectively.  This compares to -37\% and -55\% for a noon burst at this time of year and latitude.  Again, the difference between day and night bursts is 3 to 4 percent.  The greater ozone depletions for a nighttime burst as compared to a daytime burst can be explained by the fact that during darkness $\mathrm{NO_y}$ can be somewhat more easily produced by a burst and accumulated. (See the discussion above on the effects of sunlight on $\mathrm{NO_y}$ compounds.)  For example, the largest column density value for the March, equator, burst at noon is $74.23 \times 10^{16} ~\mathrm{cm^{-2}}$, while for the similar case at midnight the value is $74.52 \times 10^{16} ~\mathrm{cm^{-2}}$.

\begin{figure}
\plotone{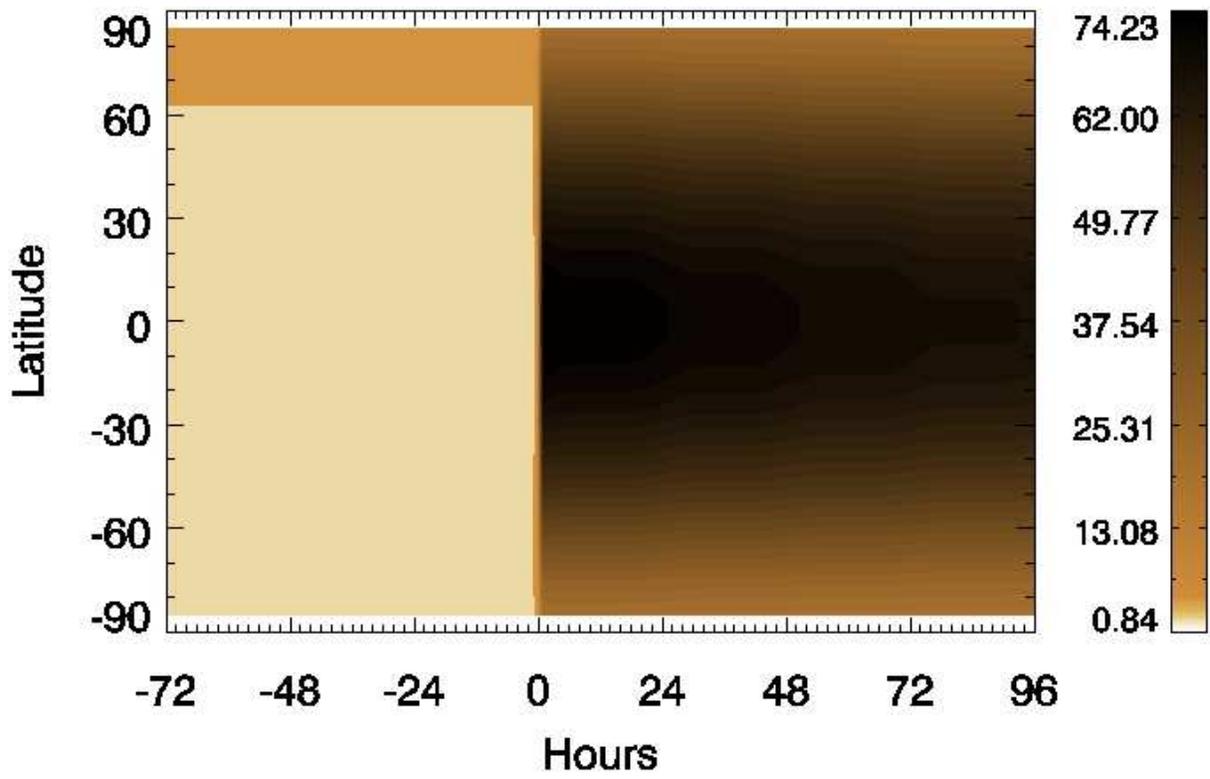}
\caption{Column density of $\mathrm{NO_y}$ in units of $10^{16}~\mathrm{cm^{-2}}$.  Hourly results for a $100~\mathrm{kJ/m^2}$ burst over the equator in late March.  (Burst occurs at time 0.  See color figure online.)
\label{fig:NOy_coldens-diurnal}}
\end{figure}

\begin{figure}
\plotone{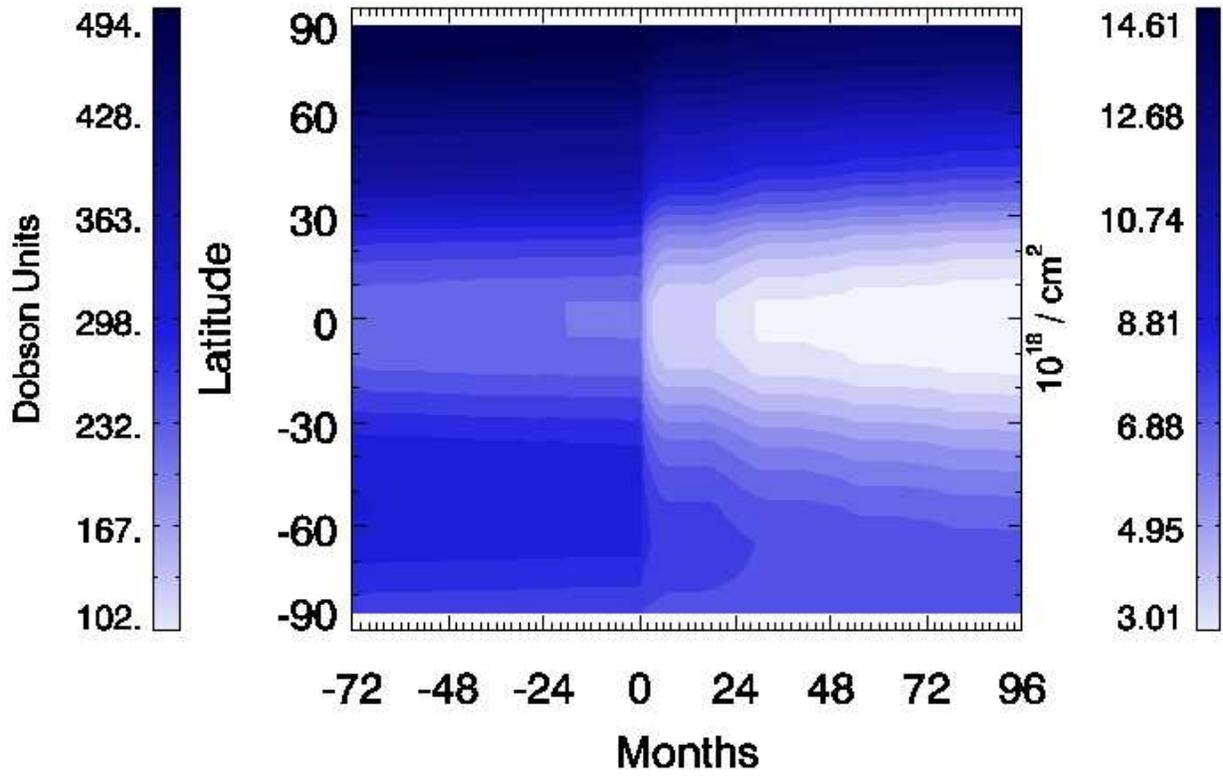}
\caption{Column density of $\mathrm{O_3}$ with scales for both Dobson units (left) and $10^{18}~\mathrm{cm^{-2}}$ (right).  Hourly results for a $100~\mathrm{kJ/m^2}$ burst over the equator in late March. (Burst occurs at time 0.  See color figure online.)
\label{fig:O3_coldens-diurnal}}
\end{figure}

The long term results for a midnight burst (i$.$e$.$ results like those in Figs.~\ref{fig:NOy_coldens}--\ref{fig:O3_perchg}) are nearly identical to those for a noon burst and so we will not reproduce them here.  We do present, however, in Figs.~\ref{fig:NOy_coldens-diurnal} and \ref{fig:O3_coldens-diurnal} the results of the short term model for $\mathrm{NO_y}$ and $\mathrm{O_3}$ column density, showing the immediate effect of the burst.  These figures are for a burst at noon, in late March, over the equator.  The run starts at noon and the burst is input 72 hours later (at time 0 in the plots).  Note that the scales here are different than in Figs.~\ref{fig:NOy_coldens} and \ref{fig:O3_coldens}, in order to give a clearer sense of the immediate production/destruction, particularly for ozone, where maximal effect is not seen for some months after the burst.

We have also investigated bursts of various durations with total energy fixed.  We find for bursts of durations between 1 s and 1000 s there is no significant difference in long term results.  The implication that total energy is more important than duration is in concordance with the fact that ozone depletion results for a SN \citep{geh03} of the same fluence are not greatly different from our results.  In that study, a fluence of approximately $15~\mathrm{kJ/m^2}$ input over 300 days resulted in maximum globally averaged ozone depletion of about 27\%.  
For a GRB delivering the same fluence over 10 s we find a maximum globally averaged ozone depletion of 16\%.  Of course, such a GRB could be 500 times farther away and produce this similar effect.

We would like to understand why total energy input is more important than duration of that input.  The timescale of our burst range (1-1000 s) is long compared to the ionization and dissociation timescales, which are nearly instantaneous.  Additionally, the input timescale is short compared to the ozone depletion timescale, which is hours to days.  To the depletion process, all bursts in this duration range are effectively impulsive and provide equal amounts of initiating compounds.

To first order, for the energy regime in which we are working, every photon interacts high in the atmosphere by gross ionization and dissocation through many encounters.  The effect of one photon is essentially independent of another and is proportional to how many ionizations and dissociations it can cause, which just depends on its total energy.  So, the overall effect is, to first order, a result of the total energy of the burst, not its duration (within the range we consider here).  It is possible that the individual photon ``hardness'' (i$.$e$.$ the spectrum) may have an effect if it is extremely different because of the atmospheric cross section differences.  This would, however, be a rather unusual GRB.  We are currently working on this aspect, as well as including effects from the GRB afterglow.

\subsection{Varying Burst Fluence}
\label{results-10kJ}
There is a wide range of possible intrinsic energies and distances from the Earth for a burst in the last billion years (see Sect.~\ref{model-grbinput} and \citet{mel04}).  Our primary event was selected as a conservative estimate of the probable largest burst in the last billion years.  We have also modeled bursts delivering two other fluence values to the Earth.  

We will first discuss the results for a burst of incident fluence $10~\mathrm{kJ/m^2}$.  This fluence corresponds to our standard burst at a distance of approximately 6.5 kpc.  We will not present as complete a set of results as that discussed above for the $100~\mathrm{kJ/m^2}$ case, since the main qualitative features due to time of year, latitude and time of day are the same, though the intensity in all cases for this lower fluence is, of course, reduced.  We present results for a burst in late March over the equator, for comparison with results discussed in \citet{thom05} for the $100~\mathrm{kJ/m^2}$ case.  We also present results for a burst in late September over $+45^\circ$ latitude.  For the $100~\mathrm{kJ/m^2}$ case, the highest ozone depletions occur for a burst in September over the north pole.  However, a burst over $+45^\circ$ is more likely than a polar burst (due to surface area considerations), so we have chosen this combination as a probable upper limit of effects for this fluence case.

\begin{figure}
\plotone{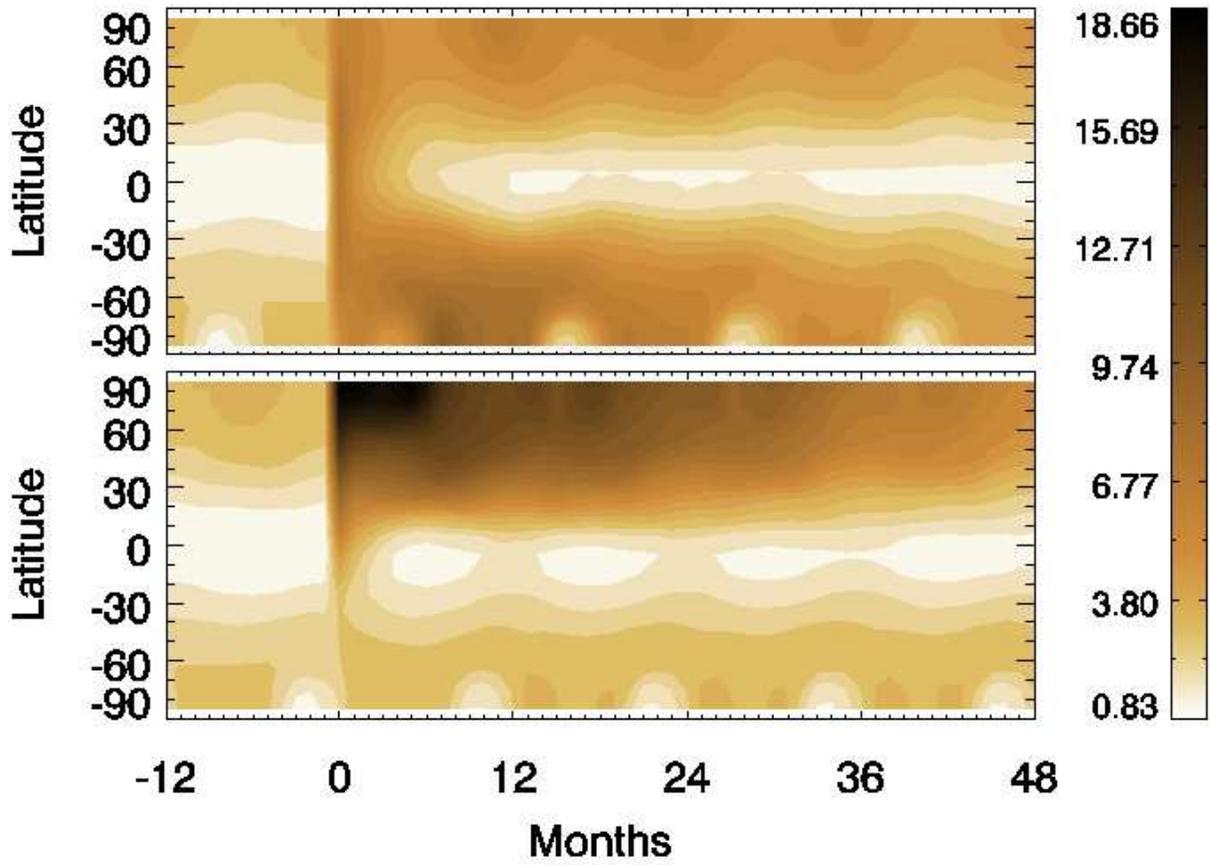}
\caption{Column density of $\mathrm{NO_y}$ in units of $10^{16}~\mathrm{cm^{-2}}$, for $10~\mathrm{kJ/m^2}$, March, Equator (top panel) and September, $+45^\circ$ (bottom panel) cases. 
(Bursts occur at month 0.)  Note that the scale here is not the same as that for Fig.~\ref{fig:NOy_coldens}. (See color figure online.)
\label{fig:NOy_coldens-10k}}
\end{figure}

\begin{figure}
\plotone{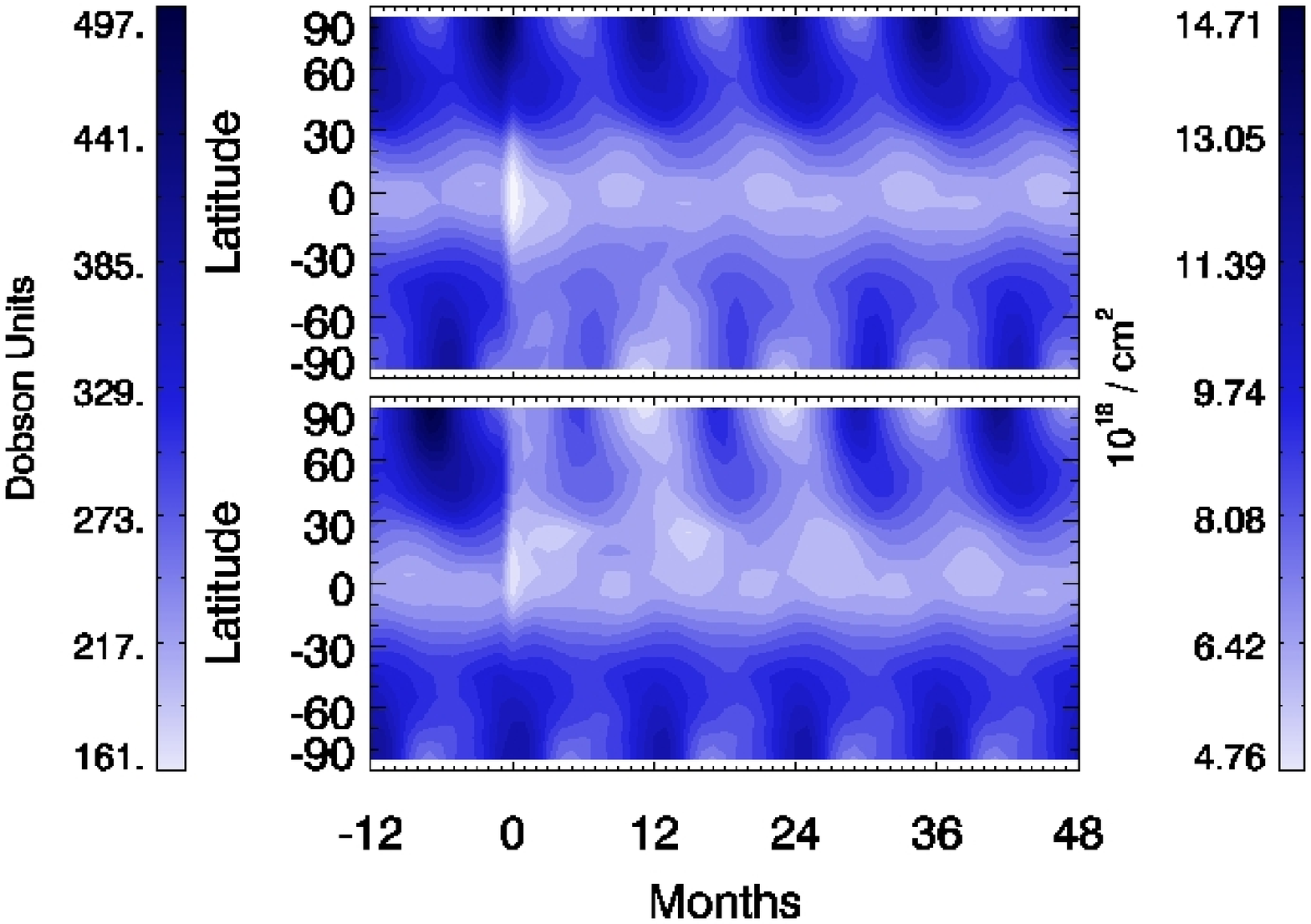}
\caption{Column density of $\mathrm{O_3}$ with scales for both Dobson units (left) and $10^{18}~\mathrm{cm^{-2}}$ (right), for $10~\mathrm{kJ/m^2}$, March, Equator (top panel) and September, $+45^\circ$ (bottom panel) cases.  (Bursts occur at month 0.)  Note that the scale here is not the same as that for Fig.~\ref{fig:O3_coldens}. (See color figure online.)
\label{fig:O3_coldens-10k}}
\end{figure}

\begin{figure}
\plotone{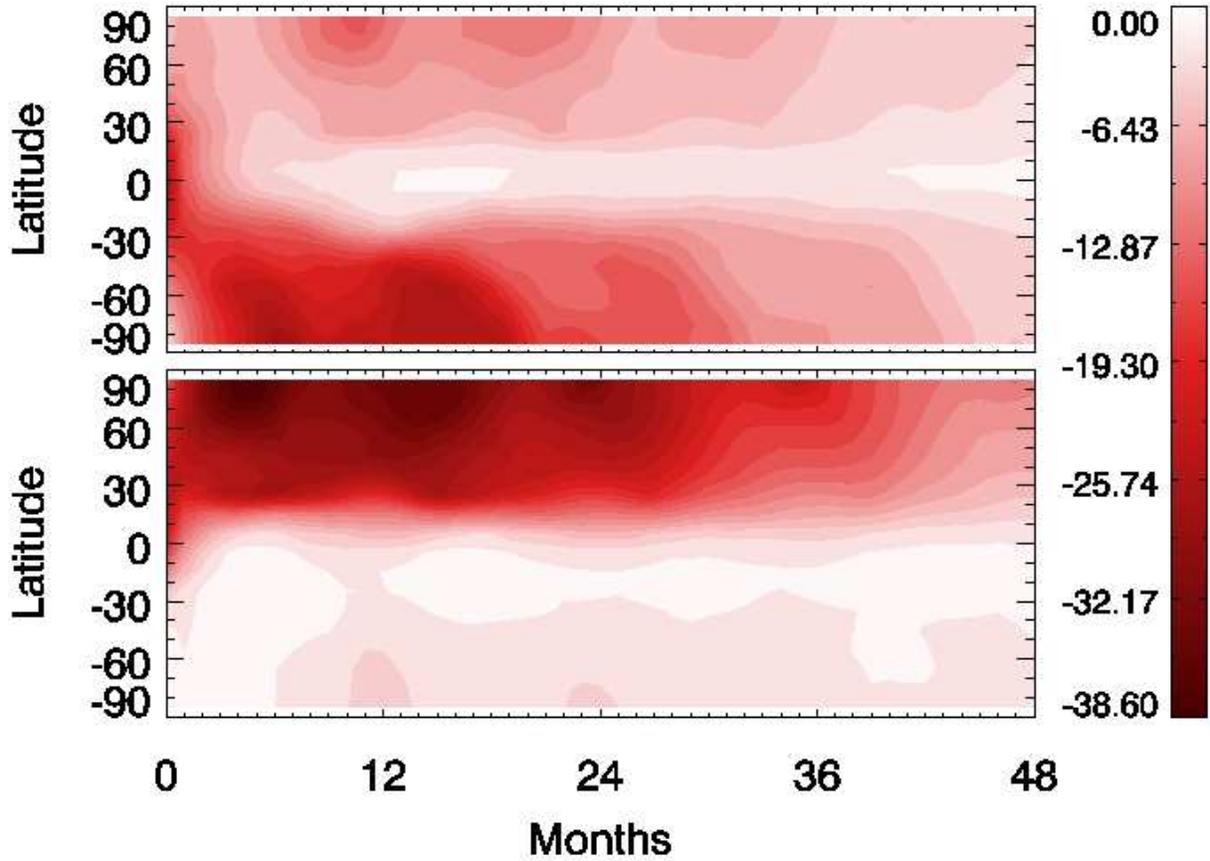}
\caption{Pointwise percent change in column density of $\mathrm{O_3}$ (comparing runs with and without burst), for $10~\mathrm{kJ/m^2}$, March, Equator (top panel) and September, $+45^\circ$ (bottom panel) cases.  Note that the scale here is not the same as that for Fig.~\ref{fig:O3_perchg}.  In contrast to Fig.~\ref{fig:O3_perchg}, there are no areas of positive percent change in these cases. (See color figure online.)
\label{fig:O3_perchg-10k}}
\end{figure}

\begin{figure}
\plotone{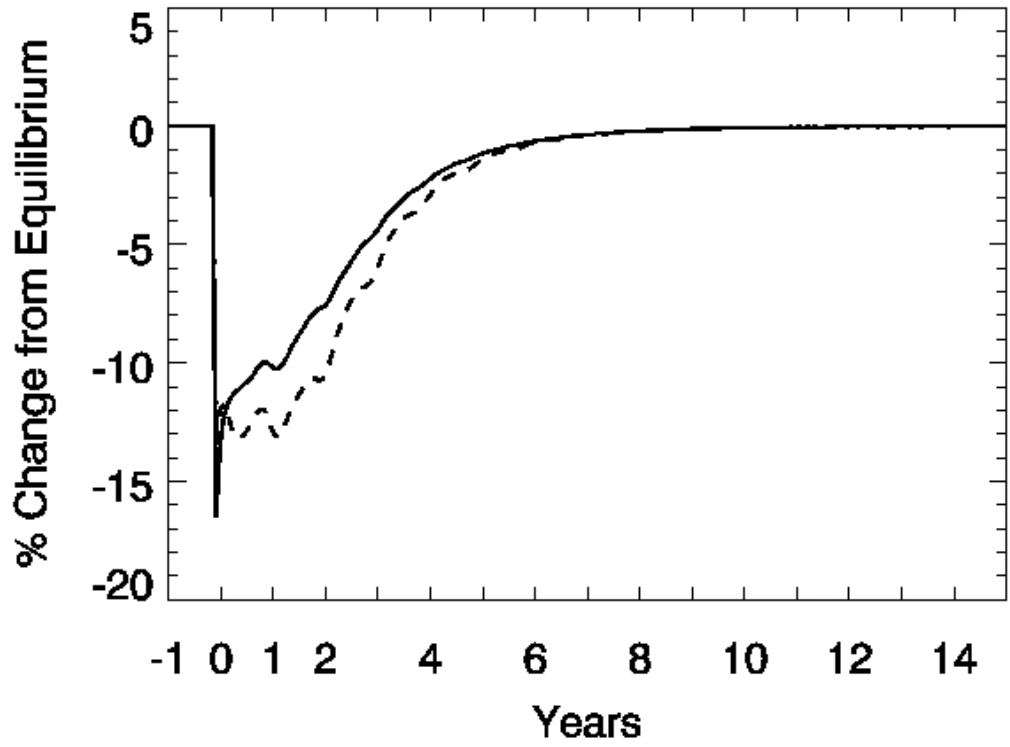} 
\caption{Percent change in globally averaged column density of $\mathrm{O_3}$, for $10~\mathrm{kJ/m^2}$, March, Equator (solid line) and September, $+45^\circ$ (dotted line) case. 
\label{fig:O3_perchg-glob-10k}}
\end{figure}

Atmospheric results for a burst of incident fluence $10~\mathrm{kJ/m^2}$ in late March over the equator, and in late September over $+45^\circ$ latitude are shown in Figs.~\ref{fig:NOy_coldens-10k} -- \ref{fig:O3_perchg-glob-10k}, which are similar to Figs.~\ref{fig:NOy_coldens} -- \ref{fig:O3_perchg-glob} for the $100~\mathrm{kJ/m^2}$ cases.
Like the previous fluence cases discussed, immediate atmospheric effects are seen in production of $\mathrm{NO_y}$ and destruction of $\mathrm{O_3}$.  We see here again that effects from a burst over $+45^\circ$ latitude are mostly contained within that hemisphere.  Likewise, greater ozone depletion is seen at the polar regions for both burst latitudes and primarily in the southern polar regions for the equatorial burst, as seen for the previous cases as well and described in more detail above.  Note here that no ozone production ``spikes'' are present, though the depletion is smaller in the same time-frame as the spikes in the previous cases.  This difference is simply due to the much smaller quantity of $\mathrm{NO_y}$ produced in this case (maximum column densities more than an order of magnitude smaller).  
Again similarly to the $100~\mathrm{kJ/m^2}$ cases, localized ozone depletion is greater and globally averaged depletion is lower for the $+45^\circ$ burst latitude, as compared to an equatorial burst.  See Table~\ref{tbl:4} for values of maximum localized and globally averaged ozone depletion.  
In comparison to the  $100~\mathrm{kJ/m^2}$ March, Equator case, this fluence yields only $0.44$ times the globally averaged ozone depletion and $0.51$ times the localized depletion.  Similarly, for the September, $+45^\circ$ burst this fluence yields only $0.39$ of the globally averaged ozone depletion and $0.52$ of the localized depletion.  
It is interesting to note that the effects as measured with these quantities are only about half as great for a fluence which is an order of magnitude smaller.  This is likely due to the fact that there tends to be a ``freeze-out'' of NO production which will limit the relative impact of higher fluence bursts.

 Figure~\ref{fig:O3_perchg-glob-10k} shows the globally averaged ozone depletion over time for both $10~\mathrm{kJ/m^2}$ cases presented.  Note that the depletion for the September, $+45^\circ$  case is longer lived (as can also be seen in Fig.~\ref{fig:O3_perchg-10k}).  Significant depletion (greater than 10\%) is evident only as long as 2 years after the burst (1 year for the March, Equator case) and full recovery is achieved by about 8 years after the burst in both cases.

Although our model is probably not capable of correctly handling the chemistry, heating and transport for a burst of much higher fluence than $100~\mathrm{kJ/m^2}$, we have also modeled a $1~\mathrm{MJ/m^2}$.  (See Sects.~\ref{uncert-NOy} and \ref{uncert-temp} for a discussion of uncertainties relevant here.)  This fluence corresponds to a burst at a distance of about 600 pc, which is the closest probable burst in the last billion years, based on less conservative assumptions, as discussed in \citet{mel04}.  Due to the high level of uncertainty in modeling this fluence, we will not present full results.  However, for comparison with the above results, our modeling indicates that a burst delivering this fluence in late March over the equator yields globally averaged ozone depletion of -65\% (1.8 times that for the $100~\mathrm{kJ/m^2}$ case) and localized depletion of -80\% (1.5 times the $100~\mathrm{kJ/m^2}$ case).  It is again interesting to note that roughly a factor of two separates these depletion values for fluences an order of magnitude different.

It is interesting to note the trends of $\mathrm{NO_y}$ production and $\mathrm{O_3}$ depletion for the three fluence cases presented, with the strong caveats of the uncertainty of our modeling of the $1~\mathrm{MJ/m^2}$ fluence and the fact that we have only three data points.  That said, we find that $\mathrm{NO_y}$ production scales linearly 
with fluence while the percent change in globally averaged column density of $\mathrm{O_3}$ scales with fluence as a power law with index $\sim 0.3$ over this fluence range.  
The slower increase in $\mathrm{O_3}$ depletion with fluence as compared with increase in $\mathrm{NO_y}$ production shows that the efficiency of depleting ozone decreases with increased $\mathrm{NO_y}$ production.  This is partly due to increased interaction of $\mathrm{NO_y}$ with other species, which limits the effectiveness both of $\mathrm{NO_y}$ on ozone as well as other species (such as Cl and Br) which deplete ozone.
There is also higher \emph{production} of  $\mathrm{O_3}$ in cases where higher levels of  $\mathrm{NO_y}$ are seen.  This effect is seen both in total column density (at polar spring, see Sect.~\ref{results-100kJ}) and at altitudes below the normal peak of ozone concentration (see Fig.~\ref{fig:O3_perchg_altlat}).  Additionally, there is only so much ozone present to deplete and complete depletion is never reached, but only approached asymptotically.  For instance, at 32 km altitude, the pointwise percent change in ozone one month after the burst is -64\% for a $10~\mathrm{kJ/m^2}$, -91\% for a $100~\mathrm{kJ/m^2}$ burst and -98\% for a $1~\mathrm{MJ/m^2}$.  
This shows that the depletion simply cannot scale linearly with fluence, even though the $\mathrm{NO_y}$ production does, since even the middle fluence value depletes nearly all of the ozone at that altitude.  

\subsection{Biological Effects} 
\label{results-bio}
\begin{figure}
\plotone{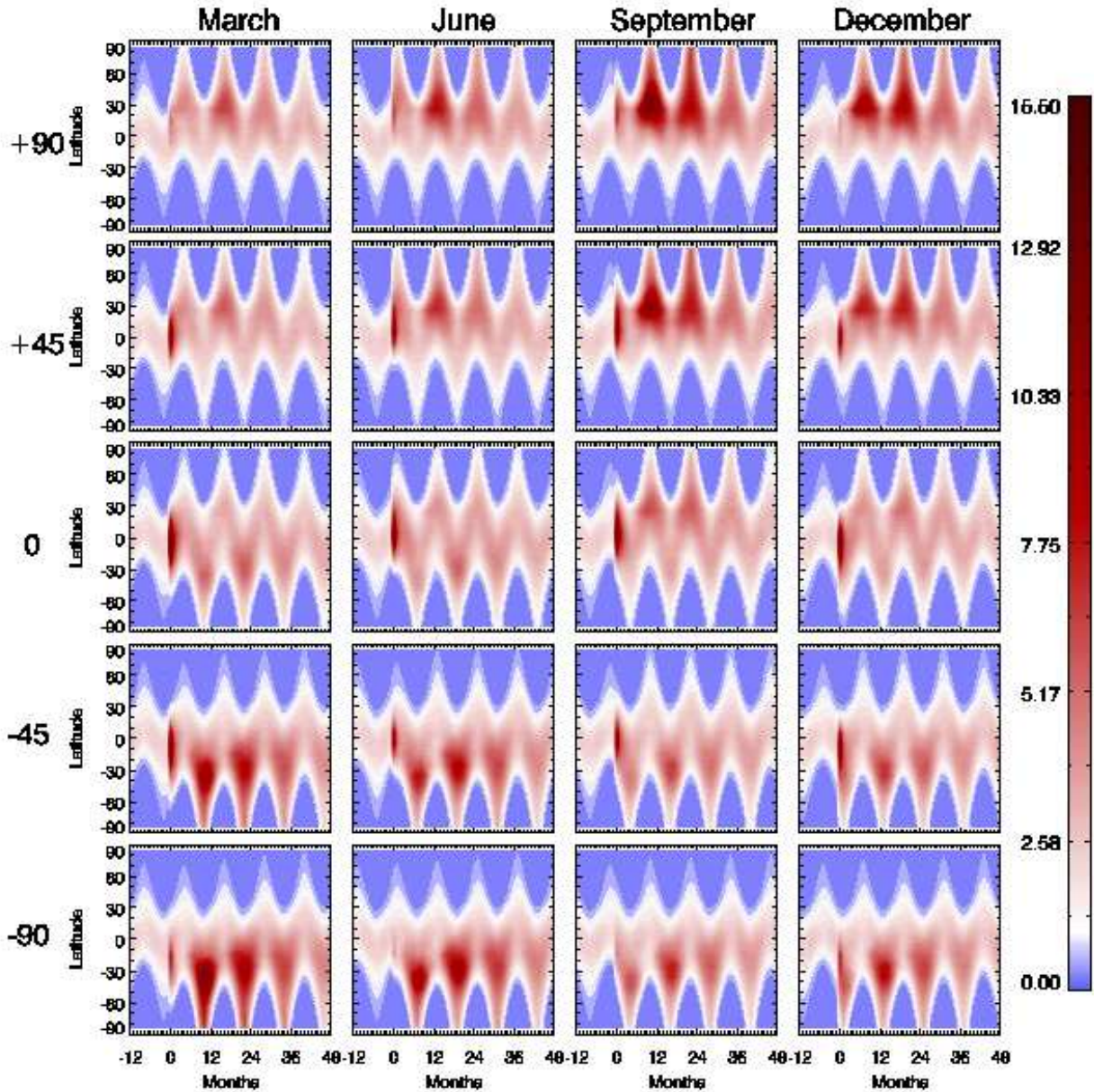}
\caption{Relative DNA damage (dimensionless), normalized by the annual global average damage in the absence of a GRB, for $100~\mathrm{kJ/m^2}$ bursts over latitudes $+90^\circ$, $+45^\circ$, the equator, $-45^\circ$, and $-90^\circ$, at the equinoxes and solstices. (Bursts occur at month 0.)  Note that white indicates values between 1.0 and 0.0 (see color figure online).  Experiments suggest significant mortality for marine microorganisms should certainly exist in areas where this measure exceeds 2.
\label{fig:dna}}
\end{figure}

\begin{figure}
\plotone{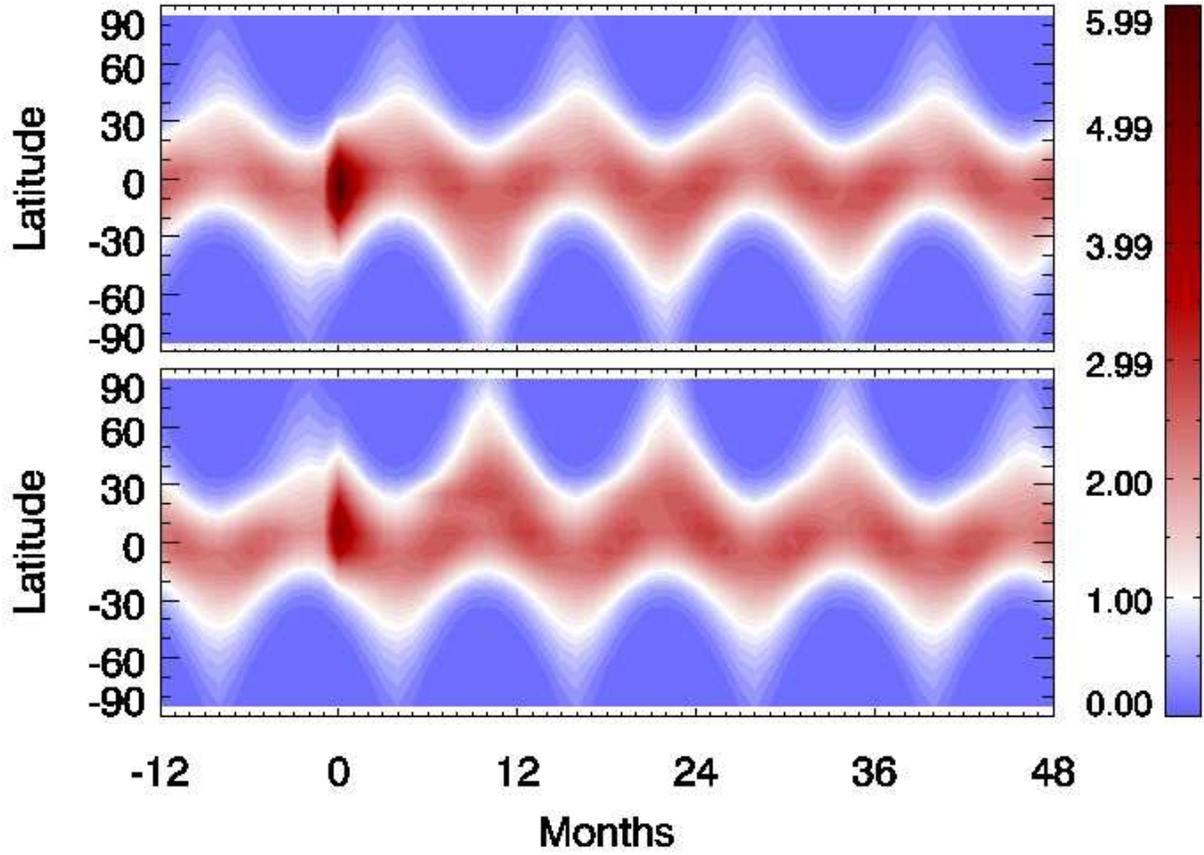}
\caption{Relative DNA damage (dimensionless), normalized by the annual global average damage in the absence of a GRB, for $10~\mathrm{kJ/m^2}$, March, Equator (top panel) and September, $+45^\circ$ (bottom panel) cases. (Bursts occur at month 0.)  Note that white indicates values between 1.0 and 0.0 (see color figure online).  Note that the scale here is not the same as that for Fig.~\ref{fig:dna}.
\label{fig:dna-10k}}
\end{figure}

Ozone normally absorbs about 90\% of solar UVB (wavelengths between 290--315 nm) radiation.  This wavelength range is particularly damaging to organisms since DNA absorbs strongly here.  Depletion of ozone due to a burst, as we have modeled, will lead to increased solar UVB reaching the surface and penetrating a few to 10's of meters in water.  
As described in Sect.~\ref{model-bio}, we have computed the direct effect of this increased UVB on life by determining the irradiance at the surface and convolving this with a biological weighting function \citep{set74,smith80} which describes the effectiveness of particular wavelengths in damaging DNA molecules (see Fig.~\ref{fig:DNAweight}).  These results are presented in Figs.~\ref{fig:dna} and \ref{fig:dna-10k}.
Other weighting functions also exist, quantifying effects such as erythema (sun burn, e$.$g$.$ \citet{mck87}), inhibition of photosynthesis (e$.$g$.$ \citet{cull92}), melanoma (e$.$g$.$ \citet{set93}), mortality (e$.$g$.$ \citet{kou99}), etC.  While these all vary from each other somewhat, the variation is small.  Figure 7.1 of \citet{jag85} shows several different weighting functions, all of which have similar shapes over our wavelength range.

Figure~\ref{fig:dna} shows DNA damage for all burst latitude and time of year cases discussed for $100~\mathrm{kJ/m^2}$ fluence (Sect.~\ref{results-100kJ}).  This includes the burst in late March over the equator, which was first presented in \citet{thom05}.
Figure~\ref{fig:dna-10k} shows DNA damage for the cases discussed for $10~\mathrm{kJ/m^2}$ fluence (Sect.~\ref{results-10kJ}).  
We have normalized the plots by dividing the damage by the annual global average damage in the absence of a GRB.  Note that the scales in the figures are not the same, in order to allow for greater visibility of features.  Also, white areas indicate relative DNA damage values between 1.0 and 0.0 (see color figures online).  Experiments suggest significant mortality for marine microorganisms should certainly exist in areas where this measure exceeds 2.

Several interesting features may be noted.  For many cases, maximum DNA damage is evident immediately around the equator and is fairly short-lived.  However, for some cases (particularly polar bursts), the maximum values occur somewhat later at mid latitudes (e$.$g$.$ around $\pm30^{\circ}$).  In the long term (more than a few months), greatest DNA damage is apparent at mid to low latitudes for all burst cases.  This heightened damage lasts several months at a time, recurring annually.  
This latitude dependence and annual recurrence is due to combination of the $\mathrm{O_3}$ depletion effects with the Sun incidence angle, length of day, etc.  Note that the highest DNA damage values occur for cases with the highest localized ozone depletion.

Like the atmospheric results, the lower fluence DNA damage results show similar qualitative properties to that for the higher fluence (greatest damage initially at the equator, long-term enhanced damage at mid latitudes, annually recurring), but less intense.  The maximum DNA damage for the $10~\mathrm{kJ/m^2}$ cases occurs with a March burst over the equator and is $0.39$ of the maximum for the $100~\mathrm{kJ/m^2}$ cases (including a March, equator burst).  Note that this is a smaller fraction than that for ozone depletion (local or globally averaged) when comparing between fluences for this burst case.  This is due to the non-linear relationship between DNA damage and ozone depletion \citep{mad98}.  Long-term enhancement of damage is relatively small and shorter lived here, indicating that damage is probably only significant for the first 2 months or so.  These features are understandable given the relatively small ozone depletions at mid latitudes for this fluence.

It is important to note that greater DNA damage is evident at mid to low latitudes, in the long term, for \emph{all} cases.  One might think that this damage would be countered by a greater evolved UVB-resistance in organisms at low latitudes.  However, at least for modern organisms, there is no evidence that temperate zone phytoplankton are any more UVB-resistant than Antarctic plankton \citep{prez04}.  Thus, one might predict that greater ecological damage and extinction would be likely near the equator. 
It is interesting that the late Ordovician mass extinction seems to be alone in having recovering fauna preferentially derived from high-latitude survivors \citep{jab04}.

\subsection{Opacity and Nitrate Deposition}
\label{results-opacity}
Two additional effects of a GRB impact, first discussed for general ionizing events by \citet{reid78}, are reduction of sunlight due to opacity in the visible of $\mathrm{NO_2}$ and nitrate deposition due to rainout of $\mathrm{NO_y}$ in the form of $\mathrm{HNO_3}$ (nitric acid rain).  Here we quantify these two effects using our modeling results for the various latitude and time of year cases discussed in Sect.~\ref{results-100kJ} for a  $\mathrm{100~kJ/m^2}$ fluence burst.  A subset of the results in this section are discussed elsewhere \citep{mel05}.

\subsubsection{Opacity Due to $\mathrm{NO_2}$}
\label{opacity-no2}
\begin{figure}
\plotone{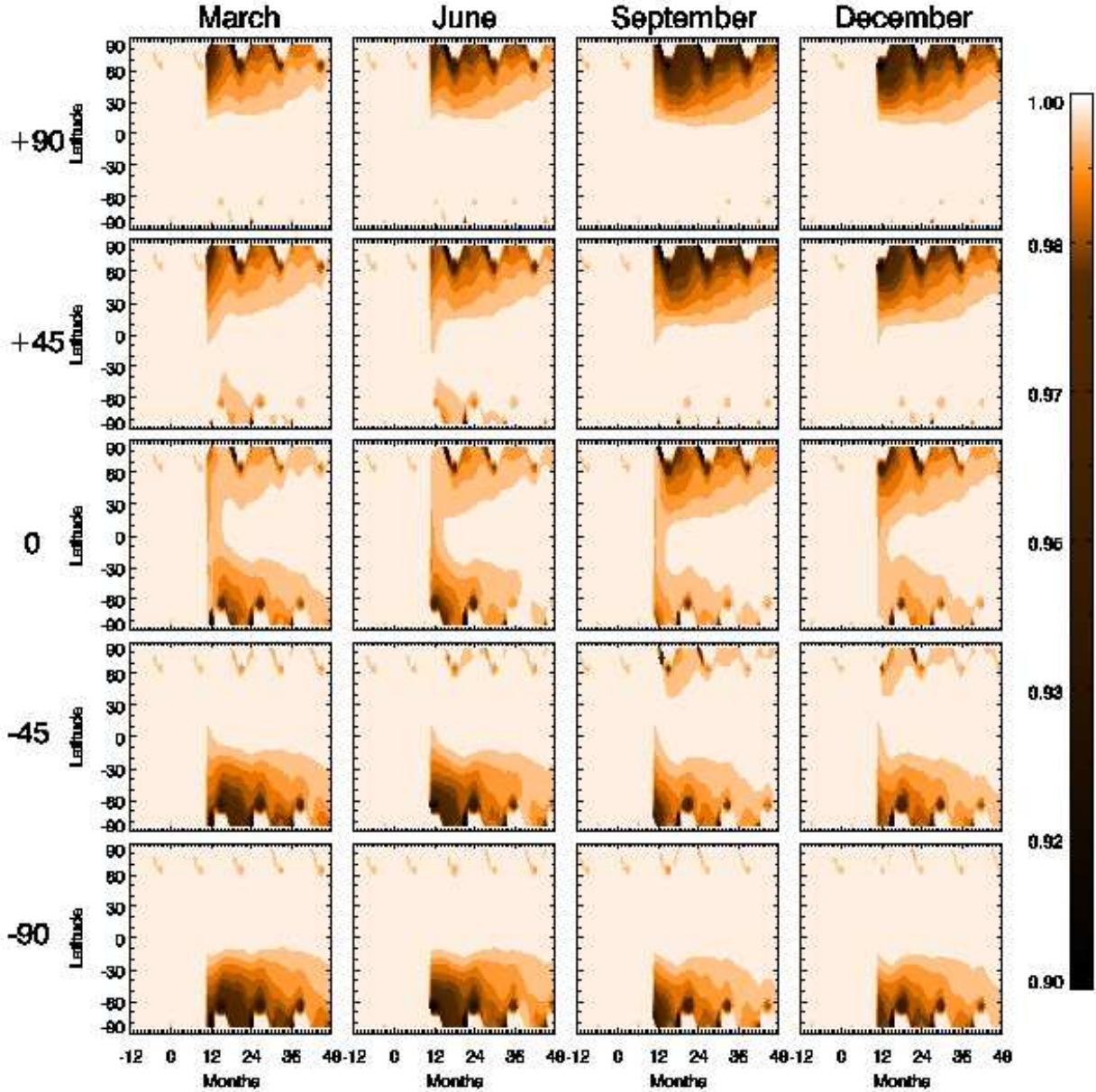}
\caption{The time development of relative solar fluence reaching the surface for $100~\mathrm{kJ/m^2}$ bursts over latitudes $+90^\circ$, $+45^\circ$, the equator, $-45^\circ$, and $-90^\circ$, at the equinoxes and solstices.
Relative fluence is the computed amount divided by that in the absence of $\mathrm{NO_2}$ absorption.  The value is set to 1 for total polar darkness.
\label{fig:opacity}}
\end{figure}

As discussed above, $\mathrm{NO_2}$ is a major compound generated by a GRB impact.  It has a major role in $\mathrm{O_3}$ depletion, but also absorbs strongly in the visible, giving it a brown cast.  Such absorption may easily lower global temperatures, if sufficient $\mathrm{NO_2}$ is formed.  In order to calculate the reduction in transparency due to $\mathrm{NO_2}$, we first find the surface irradiance at any given time in the presence of $\mathrm{NO_2}$.  This was done by convolving the solar spectral irradiance with the transparency of the Standard US Atmosphere \citep{astm99} and the transparency of the $\mathrm{NO_2}$. 
The transparency of $\mathrm{NO_2}$ is calculated using the Beer-Lambert type relation, similar to that used in this work to calculate UV irradiance, as discussed in Sect.~\ref{model-bio}.
The optical depth was computed using the absorption cross section of $\mathrm{NO_2}$ \citep{van98} and the column density through the atmosphere (computed in the model) at the angle between the Sun direction and a normal to the Earth's surface.  The surface irradiance is then integrated over wavelength and multiplied by the cosine of the angle between the Sun and the surface normal to account for insolation.  The irradiance was averaged over the course of a day.  This calculation was repeated with the $\mathrm{NO_2}$ column density set to zero to obtain the average daily irradiance in the absence of $\mathrm{NO_2}$.  The relative transparency was taken as the ratio of these two mean irradiance values, and is shown in Figure~\ref{fig:opacity}, for the burst cases discussed in Sect.~\ref{results-100kJ}.  We show a full year pre-burst, so that the change due to the burst is more obvious.

The greatest intensity and duration of reduction in sunlight reaching the surface occurs, not surprisingly, for those burst cases where the most $\mathrm{NO_y}$ column density is produced in a local area (see Fig.~\ref{fig:NOy_coldens}).  
Our results are suggestive that climate change may be possible from the several years' reduced sunlight.  For example, an estimated reduction of only 0.36\% in solar flux during the Maunder Minimum may have caused the ``Little Ice Age'' in Europe \citep{hoyt93}.
In our simulations, the greatest $\mathrm{NO_2}$ buildup takes place at high latitudes, partially persisting during polar summer, which ought to have a greater effect, reducing the melting of ice there. A glaciation accompanied the late Ordovician extinction.  Its initiation is not well understood, but may have required a perturbation such as that found here in order to trigger the instability \citep{herr03,herr04
}.
Further progress in understanding this process will require detailed climate modeling.

\subsubsection{Nitrate Deposition}
\label{opacity-no3}
$\mathrm{NO_2}$ reacts with hydroxyl ($\mathrm{OH}$, a product of dissociating water vapor) to make nitric acid, $\mathrm{HNO_3}$.  This is precipitated out in rain or snow, which is one of the primary ways the atmosphere returns to equilibrium after our model burst.  The global mean surface density of nitrate is interesting, of order the amount deposited by lightning ($6\times10^{-3}~\mathrm{g/m^2}$) over several years, or the amount used in a typical agricultural application.  Nitrogen is essential for life, but atmospheric $\mathrm{N_2}$ is nearly unavailable to most organisms due to the strength of the nitrogen triple bond.  Most biota respond strongly with increased growth rates due to nitrate deposition \citep{sch97}.  Acids would stress portions of the biosphere (e$.$g$.$ \citet{hat00}, but after immediate titration should act as fertilizer \citep{str96,shi05}.  

In order to compute nitrate deposition, we used $\mathrm{HNO_3}$ rainout data as a reasonable measure, since this is the primary pathway by which a GRB would increase nitrate deposition.  Rainout is directly computed by the atmospheric model and is empirically based.  Using both $\mathrm{HNO_3}$ concentration and rainout coefficients from the model, we computed $\mathrm{HNO_3}$ rainout rates as a function of time and latitude. Significant rainout builds up at least a few months after the burst and the peak yearly deposition occurs in the third year after the burst in all cases presented here.

Our rainout estimates, and in particular their geographic pattern, must be regarded as only exemplary, since they are based on current empirical data.  Rainout is more strongly affected than are ozone depletion and radiation transparency by the configuration of surface features, such as land versus sea, mountains, etc., and overall temperature, which would be different at other times.  These results should be viewed as a rough guide to what may be expected.  

\begin{figure}
\plotone{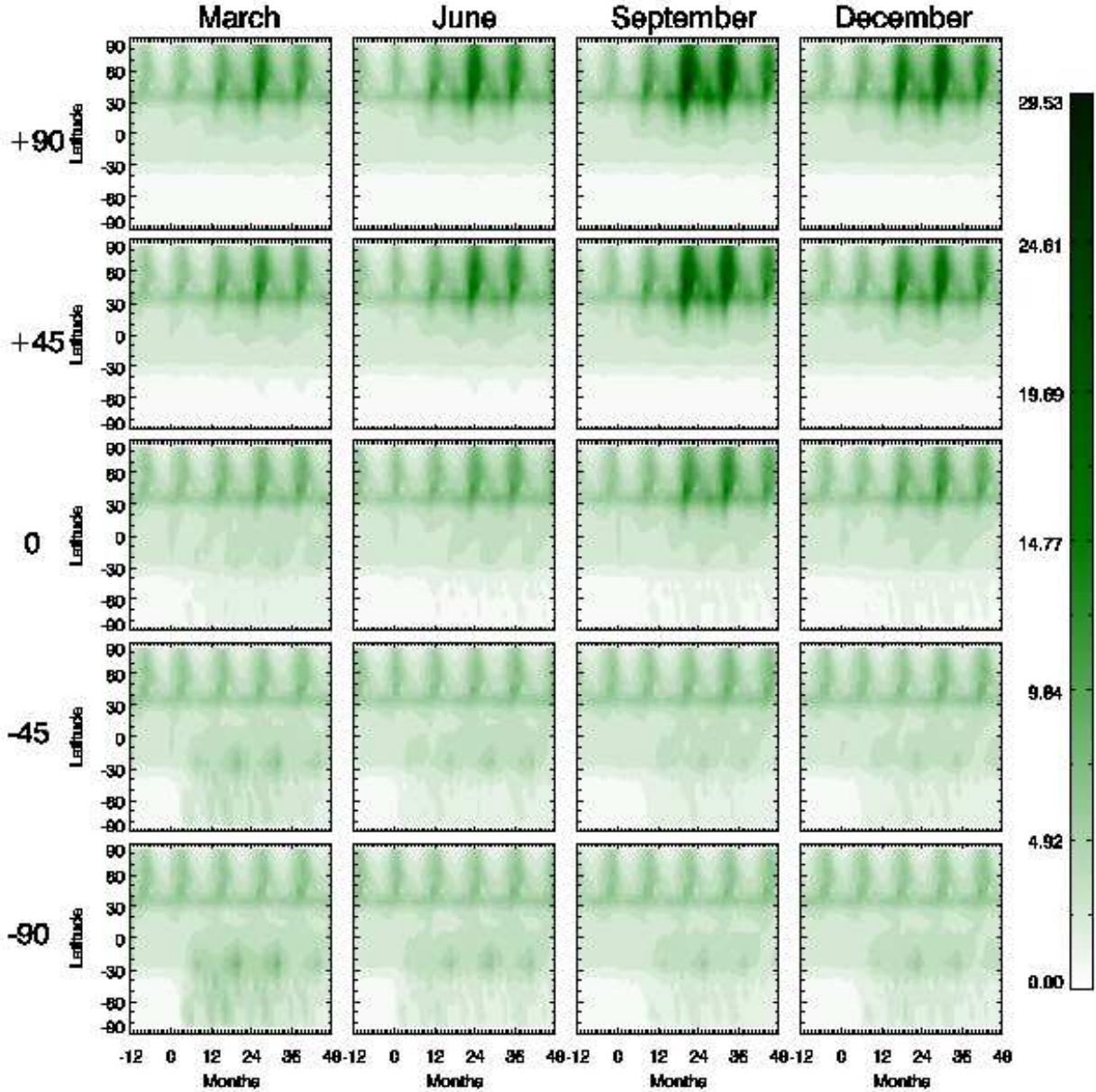}
\caption{Rainout of N as nitrate ($\mathrm{NO_3^-}$) in units of $10^{-10}~\mathrm{g/m^2/s}$ for $100~\mathrm{kJ/m^2}$ bursts over latitudes $+90^\circ$, $+45^\circ$, the equator, $-45^\circ$, and $-90^\circ$, at the equinoxes and solstices.
The pre-burst background includes all non-anthropogenic rainout at modern times, and would be somewhat lower prior to the advent of terrestrial nitrogen-fixing plants.
\label{fig:nitrates}}
\end{figure}

Figure~\ref{fig:nitrates} shows nitrate rainout rates (flux of N as nitrate) pre- and post-burst for the cases discussed in Sect.~\ref{results-100kJ} for a $\mathrm{100~kJ/m^2}$ fluence burst.  Much of the latitude asymmetry in this case is due to the fact that there is presently more precipitation in the northern hemisphere, especially north of about $45^{\circ}$ latitude, as compared to that south of $45^{\circ}$ latitude. 
This is primarily due to warmer temperatures at high northern latitudes as compared with high southern latitudes (warmer air can hold more moisture) \citep{stull00}.  More land mass at high northern latitudes also contributes, since one trigger of precipitation is uplift flow over mountains.

This latitude asymmetry would not be a feature expected at other times.  However, we believe our global averages (see Table~\ref{tbl:no3}) are reasonable guidelines to effects at earlier times.  Our baseline rainout rate is empirical and will also include a small additional component due to biological sources.  N fixation by land plants would not be a significant contribution at the Ordovician.  However, most biogenic nitrogen is directly deposited on land or in water and not from rainout.  The background rate due to lightning constitutes at most 70\% of the pre-burst background as shown in this Figure \citep{sch97}.  Thus, a burst at the Ordovician or earlier would cause a somewhat greater relative change in nitrate rainout rate.

Table~\ref{tbl:no3} lists global average annual nitrate rainout above baseline for all cases presented in Fig.~\ref{fig:nitrates}.  For reference, the baseline (background, without a burst) global average annual rainout value is $8.33 \times 10^{-3} ~\mathrm{g/m^2}$.
The values for several burst cases (especially those at $+90^{\circ}$ and $+45^{\circ}$) are similar or somewhat greater than the total generated by lightning ($6\times10^{-3}~\mathrm{g/m^2}$) and other non-biogenic sources \citep{sch97} and values for all cases are at least the same order of magnitude.  Greater values for bursts at high northern latitudes are partly due to the higher levels of rainout in the northern hemisphere in our model.  However, we also note that for a burst in September over the equator the value is close to that for lightning. 


\section{Uncertainties}
Several sources of uncertainty exist in our modeling results.  Below we discuss issues associated with our computation of ionization profiles; possible effects of redistributed gamma-ray photons; possible effects of the GRB afterglow; self-limitation of the production of $\mathrm{NO_y}$; the lack of some reactions in the model; the lack of temperature and dynamics feedback in the model.

\subsubsection{Ionization Profiles}
\label{uncert-ions}
\begin{figure}
\plotone{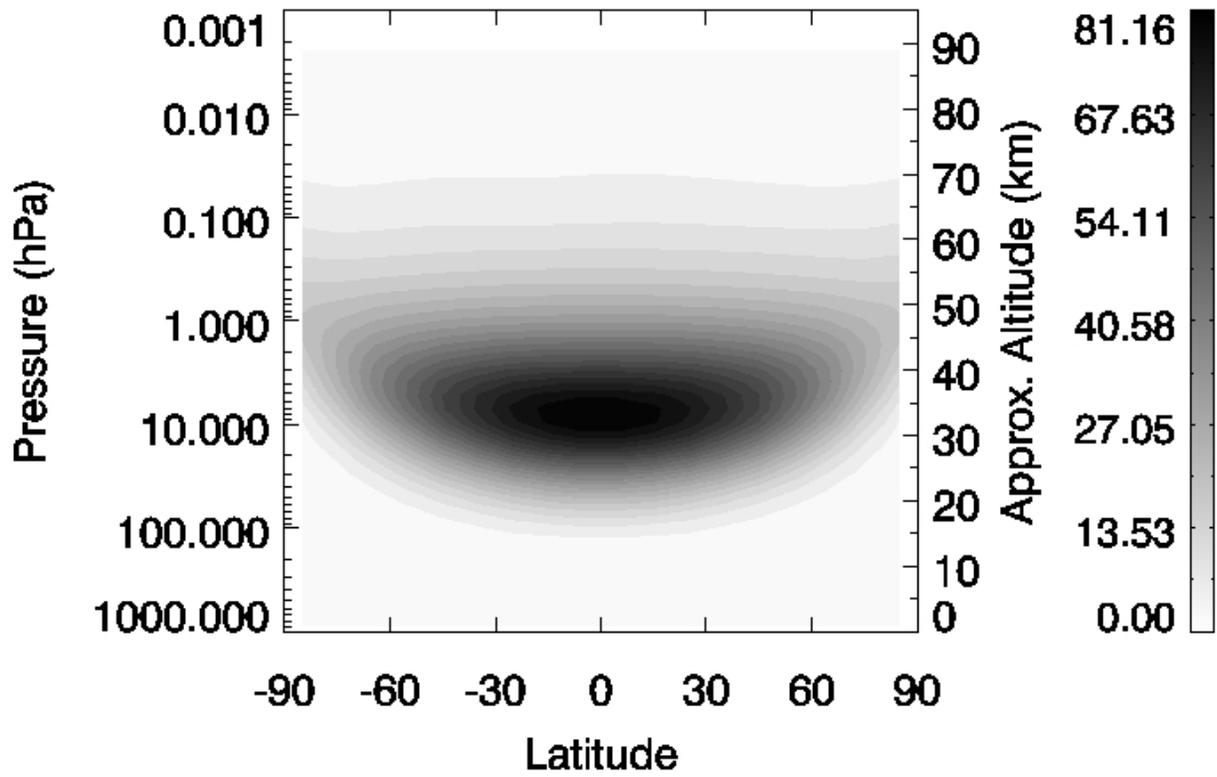}
  \caption{Energy deposition ($10^5 ~\mathrm{MeV/cm^3}$) for a $\mathrm{100~kJ/m^2}$ burst over the equator, in late March. \label{fig:ions}}
\end{figure}

Our ionization profiles are computed (as described in Sect.~\ref{model-grbinput}) using simple energy-dependent attenuation coefficients, rather than by a full radiative transfer calculation.  
We have adopted this technique from \citet{geh03}, modifying the functional form of the spectrum.  The altitude of maximum energy deposition in \citet{geh03} is around 34 km, which compares favorably with an altitude of around 32 km from a full radiative transfer method (as reported in \citet{smith04}).  In our case, as shown in Fig.~\ref{fig:ions}, we find the altitude of maximum energy deposition to be similar, around 33 km.

\subsubsection{$\mathrm{NO_y}$ Production}
\label{uncert-NOy}
Some uncertainties in the model's computation of the production of $\mathrm{NO_y}$ exist, primarily due to the lack of some reactions involving NO in the model.  The reaction $\mathrm{N + N \rightarrow N_2 + \gamma}$ is not included in the model and could limit production of NO by reducing the amount of N available to react with $\mathrm{O_2}$.  However, this reaction is unlikely due to its limited phase space and the simultaneous requirement of momentum and energy conservation.  A somewhat more likely reaction which is not included in the model is $\mathrm{N + N + M \rightarrow N_2 + M + \gamma}$, where M is an arbitrary atmospheric constituent, effectively either $\mathrm{N_2}$ or $\mathrm{O_2}$. 
This is not likely to be a large effect, since for our $\mathrm{100~kJ/m^2}$ fluence burst the N abundance compared to $\mathrm{O_2}$ or $\mathrm{N_2}$ is small, with $\mathrm{N/O_2 \sim N/N_2 \sim 10^{-6}}$.  Therefore, it appears much more likely that N would react with $\mathrm{O_2}$ rather than N in the reaction above.

There is a limiting reaction for NO production involving excited state $\mathrm{N(^2D)}$ which is not included in the model: $\mathrm{N(^2D) + NO \rightarrow N_2 + O}$.    
The destruction rate of NO with excited state $\mathrm{N(^2D)}$ is about twice that of the destruction rate with ground state $\mathrm{N(^1S)}$.  To better quantify this effect, we have performed simplified off-line computations to determine the maximum production of NO for various ratios of $\mathrm{N(^2D)}/\mathrm{N(^4S)}$.  For a $\mathrm{100~kJ/m^2}$ fluence burst, these computations (at 250 K) indicate production concentrations between $1.94 \times 10^{11}$ and $2.88 \times 10^{11} ~\mathrm{cm^{-3}}$ for ratios between $0.0$ and $1.0$.  This compares favorably with a maximum production in the model of $2.5 \times 10^{11} ~\mathrm{cm^{-3}}$.  This comparison, and the small variation of the production concentration with varying  $\mathrm{N(^2D)}/\mathrm{N(^4S)}$ ratio indicates that the exclusion of this particular reaction is unlikely to be a significant uncertainty.

Another possible limiting reaction for NO production which is not included in the model is $\mathrm{H + NO + M \rightarrow HNO + M}$.  The rate for this reaction is about $10^{-13} ~\mathrm{cm^3/s}$ compared to $10^{-11} ~\mathrm{cm^3/s}$ for the reaction 
$\mathrm{N + NO \rightarrow N_2 + O}$.  This could be significant if there is a very large concentration of H compared to N.  We do note a brief spike in the model results where H concentration exceeds N by 2-3 orders of magnitude for a few minutes and then quickly drops to 2-3 orders of magnitude smaller than N.  Hence, while there is likely a small perturbation here which we do not include, it is unlikely to be significant in the long run.

\subsubsection{Additional Radiation Effects}
\label{uncert-rad}
Our current modeling does not include the redistributed UV from the prompt GRB emission.  This radiation would be present in the stratosphere where it may affect ozone in two competing ways.  At less than 242 nm, UV will generally increase ozone by dissociating $\mathrm{O_2}$.  The two oxygen atoms then form ozone by reacting with $\mathrm{O_2}$.  UV at greater than 242 nm will generally decrease ozone because it breaks ozone apart and the oxygen atom has a chance of reacting with some other chemical rather than reacting with $\mathrm{O_2}$ to reform ozone.  Full treatment of this issue would require radiative transfer computations which are beyond the scope of the present work.  However, we have performed a simple test of  the  magnitude of such effects by increasing the solar flux used in the model by $10^4$ for 10 s to simulate the redistributed UV.  The value $10^4$ was chosen based on conversations with J. Scalo who has, along with D. Smith and J.C. Wheeler, performed radiative transfer computations for cases similar to ours \citet{smith04}.  The resulting long-term maximum ozone depletion is a few percent greater (42\% globally averaged), compared to a case without this increased solar flux (36\% globally averaged).  This test is not definitive since we have not attempted to accurately model the spectrum of the redistributed UV.  The results at least indicate that our results may be somewhat conservative.

We have not included in our modeling radiation from the afterglow of the GRB.  While not strictly an uncertainty in the modeling itself, inclusion of afterglow radiation could affect our results.  The ionizing fluence in afterglow X-ray may be comparable to that of the prompt gamma-rays, or somewhat less \citep{vand03,marsh03}.  While the energy of the photons is lower ($0.5-2 \mathrm{keV}$), this radiation would arrive over a longer time period (days).   This would likely lead to an modest increase in the $\mathrm{NO_y}$ production and $\mathrm{O_3}$ depletion, making our current atmospheric results somewhat conservative at worst.  A perhaps more significant effect is that much of the radiation would arrive after the ozone layer has been depleted, at least at some latitudes.  This would lead to an enhanced UVB flux at the ground, from redistributed X-ray as well as direct UV from the GRB afterglow, which would arrive over days to weeks.  This would augment the solar UVB, increasing our DNA damage values.

We have also neglected any variability of the parameter values used in the Band spectrum (see Sect.~\ref{model-grbinput}) or temporal variability of the burst.  This would have some effect on our ionization profiles which are input to the atmospheric model.  We cannot say for certain how large of an effect this would be, but we expect it to be small, partly due to the fact that our results are not significantly affected by spreading the total energy input over durations ranging from 1 s to 1000 s.  So far, the total energy appears to be the most important factor for long term results.

Some bursts have values of $\beta > -2$ in the Band spectrum, in which case $\mathrm{E_0}$ is not a peak of the energy distribution.  In these cases more photons are at higher energies.  We did not include this consideration in our modeling.  Long term ozone depletion could be slightly enhanced in these cases due to more high energy photons (and so more ionization), but if the total energy input is kept the same then we do not expect a large change to our results.

\subsubsection{Temperature and Dynamics Feedback}
\label{uncert-temp}
The GSFC atmospheric model does not include feedback on temperatures (and thereby, dynamics) due to changes in constituent values.  This will introduce some uncertainty to our results, since depleting ozone changes the energy absorption properties of portions of the atmosphere (particularly the stratosphere).  
This will have complicated effects.  Loss of ozone leads to cooling which slows depletion reactions involving $\mathrm{NO_y}$.  However, in the polar regions cooling will enhance formation of polar stratospheric clouds which increase ozone depletion (see Sect.~\ref{results-100kJ}).  \citet{ros02} have investigated the impact on ozone depletion by $\mathrm{CO_2}$-mediated stratospheric cooling. For temperature changes of about 5 K they find a few percent difference in ozone levels with and without the cooling effect.  They also find that effects are highly dependent upon latitude and time of year.

While it is difficult to know how our results may be affected, some estimates may be made.  Using a 1D temperature-constituent coupled model we have determined that changes in stratospheric temperature of order 10 K are probable.  
Key ozone depletion reactions may see rate reductions of 30 - 50\%.  This does not simply translate into changes in ozone depletion however, since there is feedback between ozone levels and temperatures, complicating the situation.  To get some sense of the effect on ozone levels due to a temperature reduction of 10 K, we have run an extreme case where all temperatures everywhere in the model are reduced by this amount.  This run (in March at the equator, for a $100~\mathrm{kJ/m^2}$ fluence burst) results in maximum localized ozone depletions which are a few (2-3) percent less than for a comparable run without the temperature change.  Therfore, we conclude that the impact of such a temperature change on our primary results (ozone depletion) is not likely to be significant.

Another effect of lack of temperature feedback is on dynamics (transport) in the model.  Temperatures (and gradients in temperature) affect pressures (and gradients in pressure) which drive dynamics.  Simple calculations indicate that pressure changes due to a 10 K temperature change are relatively minor; at least two orders of magnitude smaller than those associated with major weather systems such as tornadoes.  However, due to the complications of gradients and heating rates, it is not possible to estimate the full effect accurately.  The primary impact on our results of such uncertainties is likely to be changes in how rapidly $\mathrm{NO_y}$ compounds are dispersed through the atmosphere.  This could affect the latitude and time distribution of ozone depletion, even if total depletion levels remain similar.

\subsubsection{UVB Calculations}
\label{uncert-UV}
We have not used a radiative transfer calculation to determine the solar UVB flux at the Earth's surface, for use in computing DNA damage.  As discussed in Sect.~\ref{model-bio}, we use the Beer-Lambert relation and only consider effects of ozone absorption, neglecting scattering effects.  We believe this is justified for several reasons.  First, according to \citet{mad93}, in the UV band atmospheric absorption (primarily by $\mathrm{O_3}$) is dominant over scattering due primarily to the relative cross sections for these processes.  (The attenuation of a beam by $\mathrm{O_3}$ absorption is approximately two orders of magnitude greater than by scattering, neglecting diffuse background.)  In addition, while absorption leads to loss of energy, scattering primarily changes direction, not energy, leading to a diffuse background of similar flux as would occur without scattering.

We have checked our UVB irradiance results for the unperturbed, equilibrium atmosphere in comparison with measured values.  The typical UVB irradiance at noon in summer at the equator today is around $2-4 ~\mathrm{W/m^2}$.  Using the 1976 U.S. Standard Atmosphere ozone column density of $8.1 \times 10^{18}~\mathrm{cm^{-2}}$ in our code at noon at the summer solstice at the equator yields a result of about $3 ~\mathrm{W/m^2}$, in good agreement with observations.  Therefore, we feel confident that neglecting scattering here does not introduce significant errors.

\subsubsection{Application to Other Geologic Eras}
\label{uncert-eras}
 As discussed in Appendix~\ref{app:model}, the GSFC atmospheric model is tuned to the present day atmosphere, especially in regard to temperatures and dynamics.  Certain aspects of our results, such as greater ozone depletion in the southern polar regions and higher rainout values in the northern hemisphere, are largely dependent upon continental arrangement.  Thus, features such as a strong south polar vortex and enhanced PSC formation in that region may not be accurate for other periods in the Earth's history, when continental arrangements were not the same as today.  The latitude dependence of our nitrate deposition results are particularly suspect in application to different time periods.

We do, however, note that latitude dependence of ozone depletion is much more strongly affected by differences in what time of year a burst occurs at, as well as over what latitude (see Fig.~\ref{fig:O3_perchg}).  In particular, the North-South asymmetry in ozone depletion which appears for a March, equatorial burst is almost completely reversed for a burst over the same latitude in September.  Since the time of year and latitude of a burst are essentially random, we feel reasonably confident that at least our ozone results (as well as DNA damage results, which depend upon ozone levels) may be applied to eras when the continental configuration differs from today.  The random variability of burst latitude and season will overwhelm any systematic error due to geography.

Differences in atmospheric composition over the last billion years could present a problem for applying our results.  Oxygen levels are the most important factor for our modeling since significant $\mathrm{O_2}$ must be present in the atmosphere in order to have an ozone shield.  The oxygen level over the last billion years was generally close to today, except during the Permian period, and a significant level of ozone should have been in place over this time span as well \citep{ber03,graed93}.  Carbon dioxide levels have varied more greatly and were probably higher at earlier times \citet{kast93,ber01}, but this is not an important factor for our results.

Differences in temperatures between today and other eras may introduce more serious problems for applying our modeling.  Differences in atmospheric temperatures at different latitudes affect transport and so may change how quickly and to what extent $\mathrm{NO_y}$ compounds produced by the burst are transported from the latitude over which the burst occurs to other locations.  This effect would depend especially on temperature differences between the poles and tropics, as this affects transport via the ``cells'' of wind flow in the atmosphere.  We view our modeling as a ``best approximation'' in this regard, especially since determining atmospheric temperature profiles for past eras is difficult or impossible.

\section{Discussion}
It has been suggested \citep{mel04} that a GRB may have initiated the second largest mass extinction in the Earth's history, that of Late Ordovician period (approximately 443 My ago).  Patterns of extinction at this time are compatible with heightened solar UV radiation.  Such patterns include dependence on how deep an organism lived in the water column and how long an organism spent in a surface-dwelling planktonic larvae state \citep{chat89,bren95,shee01,bren03}.  In addition, surviving fauna which repopulated the globe after the extinction event appear to have come from areas of deeper water and higher latitude \citep{melc91,shee01,jab04}.  Our DNA damage results (Sect.~\ref{results-bio}) indicate that a GRB would most likely affect mid to low latitudes and would tend to leave survivors preferentially at higher latitudes.

The Ordovician extinction is associated with a glaciation event, surrounded by periods of stable warm climate.  Paleoclimate models \citep{herr03,herr04
} have indicated that glaciation at this time is not likely to have occurred without some external forcing mechanism. Given that even a few 10ths of a percent decrease in sunlight reaching the Earth's surface can have an effect on global temperatures \citep{hoyt93}, it seems likely that the increase in $\mathrm{NO_2}$ due to a GRB  (Sect.~\ref{opacity-no2}) could have provided the necessary perturbation to trigger this climate change.

From our nitrate rainout results (Sect.~\ref{opacity-no3}) we conclude that a burst could supplement usual sources of nitrate, and give a modest boost to the transition of plants to land \citep{gen01}, which was just beginning at the time of the Late Ordovician.
Biota are extremely responsive to increased nitrate flux \citep{str96,sch97,shi05}.  Increased productivity is indicated by isotopic excursions in $\delta ^{13}\mathrm{C}$ at the late Ordovician extinction, coincident with $\delta ^{18}\mathrm{O}$ indicating lower temperatures \citep{pat97,shee01,bren03,panc03}.
Increased productivity could lower atmospheric $\mathrm{CO_2}$, providing further positive feedback toward glaciation.

While nitrate deposition could aid land plants, it might at the same time provide further stress to some aquatic animals, especially in combination with increased UVB flux.  \citet{hat00} found significantly reduced survival and activity levels of larval frogs in the presence of a combination of increased UVB, nitrate and low pH.  This combination of stressors could occur, at least temporarily, in shallow water environments in the months following a burst, contributing to initiation of the mass extinction.

We have considered biological effects in the context of mass extinction.  However, less energetic bursts (or those at greater distances) could also have an interesting impact on mutation rates, and thereby on evolution.  \citet{sw02} have considered this effect due to the prompt flash of redistributed UV at the surface.  Our DNA damage results indicate that effects due to increased solar UVB may be just as important.  This effect may actually be more important, since the duration is much longer (months versus seconds).

We have not considered the effects of any cosmic rays that may accompany a GRB.  These would have further effects on atmospheric chemistry (including ozone depletion), would produce radioactive nuclides, and also have more direct effects on life through the production of showers of penetrating muons (e$.$g$.$ \citet{th95,dar98}).  

We are currently conducting further studies of the effects of including the GRB afterglow and varying the spectral parameters used.  A related question is the importance of X-ray rich GRBs ($\mathrm{E_0} \sim 10 ~\mathrm{keV}$) and X-ray flashes ($\mathrm{E_0} \sim 1 ~\mathrm{keV}$) (see, for instance, \citet{zha04,sak04,but05}.  It is likely that these are GRBs observed slightly off-axis.  Their galactic population may be larger than that of GRBs, but they are less luminous.  Hence, these events may present yet another source of astrobiologically interesting radiation on a scale similar to GRBs and supernovae.

As our results indicate, a GRB within 2 kpc of the Earth could cause significant damage to the biosphere through ozone depletion, and may have effects on climate through the opacity of $\mathrm{NO_2}$.  The ``shot'' of nitrates that would likely be deposited after a burst could both aid land plants in gaining ground and possibly add additional stress to aquatic organisms.  Such a burst is likely to have occurred within the last billion years and therefore should be considered, along with SNe, impacts and other events as contributors to the development of life (through both extinctions and enhancements of the mutation rate), both on Earth and on other Earth-like planets which may exist elsewhere.

\acknowledgments
A.L.M. wishes to thank W. Schlesinger, J. Scalo, J.C. Wheeler, and D. Smith for useful conversations.
B.C.T. wishes to thank J.E. Rosenfield for assistance with the temperature-constituent coupled model tests mentioned in Sect.~\ref{uncert-temp}.
B.C.T. acknowledges support from a Dissertation Fellowship from the University of Kansas.
B.C.T., A.L.M., C.M.L., and M.V.M. acknowledge support from NASA Astrobiology grant NNG04GM41G.
B.C.T. and A.L.M. acknowledge supercomputing support from the National Center for Supercomputing Applications.
D.P.H. and L.M.E. acknowledge support from Undergraduate Research Awards through the Honors Program at the University of Kansas.

This work has been submitted by B.C.T. in partial fulfillment of the requirements for the PhD at the University of Kansas.

\appendix

\section{Description of Atmospheric Model}
\label{app:model}
For this study we have utilized the NASA Goddard Space Flight Center two dimensional photochemical transport model, created and updated by Charles Jackman, Anne Douglass, Richard Stolarski, Eric Fleming, and David Considine.  The model has been described in several publications, including \citet{djs89,jack90,cdj94,jack96,flem99} and \citet{jack01}.  The following description draws upon these articles as well as an unpublished overview of the model by Francis Vitt (1994).

The model contains 65 chemical species (see Table~\ref{tbl:species}).  A ``family'' approach is used for the transport of most species.  Families include, for instance, $\mathrm{O_x}$ ($\mathrm{O_3}$, O, $\mathrm{O(^1D)}$), $\mathrm{NO_y}$ (N, NO, $\mathrm{NO_2}$, $\mathrm{NO_3}$, $\mathrm{N_2O_5}$, $\mathrm{HNO_3}$, $\mathrm{HO_2NO_2}$, $\mathrm{ClONO_2}$, $\mathrm{BrONO_2}$), $\mathrm{HO_x}$ (H, OH, $\mathrm{HO_2}$), $\mathrm{Cl_y}$ (Cl, ClO, HCl, HOCl, $\mathrm{ClONO_2}$), etc.  It is assumed that family member species are in photochemical equilibrium with each other since reactions producing interchanges among members are fast compared to photochemical and transport processes that produce net change in the family concentration.  Several other species are transported separately.  See Table~\ref{tbl:trans-spc} for a list of the transported families and species.

Listed in Table~\ref{tbl:BCs} are lower (i$.$e$.$ ground level) boundary conditions for the transported species in the model.
Species which are solely anthropogenic have been set to zero (both boundary conditions and number densities at all altitudes and latitudes).  These include CFC-11, -12, -22, -113, -114, and -115; Halon-1211, -1301, and -2402; HCFC-141b, -142b, and -123.  This was done because we are primarily interested in applying our results to pre-industrial time periods.

Some species included have natural as well as anthropogenic sources (such as $\mathrm{N_2O}$, $\mathrm{CO_2}$, $\mathrm{CH_4}$, $\mathrm{CH_3Cl}$, $\mathrm{CH_3Br}$).  We have set these to pre-industrial values (see Table~\ref{tbl:BCs}).  These compounds, especially $\mathrm{CO_2}$, $\mathrm{CH_4}$ and $\mathrm{N_2O}$ have probably varied by as much as an order of magnitude or more over the last billion years, with $\mathrm{CO_2}$ generally decreasing, $\mathrm{CH_4}$ fluctuating and $\mathrm{N_2O}$ increasing (see \citet{kast93} and \citet{ber01}).  The oxygen level over the last billion years was generally close to today, except during the Permian period, and a significant level of ozone should have been in place over this time span as well \citep{ber03,graed93}


\subsection{Species Continuity Equation}
\label{model-detail-spec}
The main calculation of the model is the solution of the longitudinally averaged species continuity equation:
\begin{equation} \label{eq-cont}
{{\partial n_i} \over {\partial t} }= -\vec{v} \cdot \nabla n_i - D(n_i) + S(n_i)
\end{equation}
where $n_i$ is the species concentration (number density) and $\vec v$ is the wind velocity vector.  The first term on the right side is advection by circulation.  $D(n_i)$ is the diffusion term, representing small-scale mixing and also effects of planetary 
and gravity waves.  
$S(n_i)$ is the photochemical source term, which is equal to the difference in production and loss terms and is dependent upon reaction rate constants, photolysis rates and other species concentrations.

Under the assumption that changes in species concentration for one time step are small, the photochemical terms in the continuity equation can be evaluated using the previous time step's values. 
The species continuity equations are solved by a process splitting method.  That is, it is assumed that the continuity equation can be separated into a product of operators \citep{mcr82}.  It is assumed that advection of species can be separated from photochemistry and diffusion.  Here, advection operates on the concentration field before diffusion and photochemistry.  

The source term, $S(n_i)$, is written
\begin{equation}
S(n_i) = P - L(n_i) 
\end{equation}
where $P$ is the sum of all production sources for species $i$ and $L$ is the loss frequency.  For species in which the loss process is not linear, $L$ is represented by a Taylor series.  The continuity equation \ref{eq-cont} then is written as a finite difference equation:
\begin{equation}
{{n^t - n^{t-1}} \over {\Delta t}} = - {{\tilde{n}- n^{t-1}} \over {\Delta t}} + P - L(\tilde{n}) + \tilde{n}{\partial L \over \partial n} \vert _{n=\tilde{n}} - n^t{\partial L \over \partial n} \vert _{n=\tilde{n}} - D(\tilde{n})
\end{equation}
where the subscript $i$ has been dropped and $n^t$ is the result, $n^{t-1}$ is the initial condition (from the previous time step), and $\tilde{n}$ is the advected field.

\subsection{Transport}
\label{model-detail-trans}
The advection scheme includes vertical and meridional (horizontal, north-south) winds.  These winds are climatological in nature and based on empirical data sets.  The primary components in deriving the winds are temperature data from the National Centers for Environmental Prediction (NCEP) and heating rates from climatological distributions of temperature, ozone, and water vapor.  Temperatures and heating rates are updated daily in the model and winds are calculated at that time.  

The advection scheme is described in detail in \citet{flem99}.  It is mass conserving and uses an upstream piecewise parabolic method of solution \citep{col84,carp90}.  A time step of 12 hours is used for advection (while the time step for constituent changes due to photochemistry is 1 day).
The coefficients of the stream function equation depend on zonal mean temperature which is based on the 17 year average (1979-1995) NCEP temperature data for the troposphere and stratosphere and the CIRA-86 empirical reference model for the mesosphere above 1 hPa (about 47 km altitude) \citep{flem90}.  Outside the tropics, winds are derived from temperature using the gradient wind relation with measurements from the UARS satellite for tropical zonal winds ($20^\circ$S -- $20^\circ$N).  Small scale horizontal mixing (eddy diffusion) coefficients are calculated from mechanical forcing due to gravity waves in the mesosphere and upper stratosphere and from the vertical temperature gradient in the troposphere and lower stratosphere.

\subsection{Photolysis and Chemical Reactions}
\label{model-detail-chem}
The model's binary and tertiary gas-phase-only reactions are listed in Tables~\ref{tbl:rxns1} and \ref{tbl:rxns2}.  Photolysis reactions are listed in Table~\ref{tbl:rxns3}.
Rates and photolysis cross sections are taken from the Jet Propulsion Laboratory (JPL) 2000 recommendations \citep{sand00}.  Reaction rates are updated daily along with temperatures.  There are 39 wavelength intervals used in the radiative transfer scheme.  Multiple scattering is applied and daytime averaged photolysis rates are computed every 10 days using a two-stream radiative transfer method \citep{her79}.  Solar photon fluxes for the wavelength intervals at the top of the atmosphere are taken from 
measurements made by the Solar Stellar Irradiance Comparison Experiment (SOLSTICE) and Solar Spectra Irradiance Monitor (SUSIM) instruments on the Upper Atmosphere Research Satellite (UARS) \citep{jack96}, 
and are reproduced here in Table~\ref{tbl:bands1}.  

In the long term version of the model (where daily averages are used), diurnal behavior of individual species must be taken into account to calculate photochemical production and loss.  Species which disappear quickly after sunset, such as O($^3P$), O($^1D$), N, NO, Cl, H, OH and $\mathrm{HO_2}$, are assumed to be zero at night.  Daytime average values of $\mathrm{NO_2}$ and ClO are are needed to calculate odd oxygen loss and these values depend upon the concentrations of $\mathrm{N_2O_5}$ and $\mathrm{ClONO_2}$ which build up at night and decrease during the day.  Sunset values of these are calculated from daytime average values and nighttime production at the expense of $\mathrm{NO_2}$ is calculated at each grid point for each time step.

Due to the importance of diurnal behavior, and the short duration of a GRB, we have used the short-term version of the atmospheric model (with time step 1 s) to compute the chemistry for several days around the time of the burst.  Diurnal behavior is important primarily because several constituents important for our study are strongly affected by photolysis (see Sect.~\ref{results-sun} for more discussion of this).

Important reactions for our purposes in this work occur on timescales of a few to tens of seconds.  For instance, in the stratosphere NO creation ($\mathrm{N + O_2 \rightarrow NO + O}$) takes about 5 s, NO destruction ($\mathrm{N + NO \rightarrow N_2 + O}$) takes about 30 s, and $\mathrm{O_3}$ depletion by NO ($\mathrm{NO + O_3 \rightarrow NO_2 + O_2}$) takes about 85 s.  Therefore, our 1 s time step seems adequate to accurately model these reactions around the time of the burst.  It should be noted that destruction of NO proceeds faster with increased concentrations of NO, down to a timescale of around a second at the maximum values seen in our results.  There is hence a feedback between creation and destruction of NO, with more creation leading to faster destruction.  This will limit the amount of NO created overall and thereby limit ozone depletion as well.  However, normal diurnal variations of NO (and $\mathrm{NO_2}$, which also acts to deplete ozone) have a strong effect on the total amounts of these compounds which are present at any one time (see Sect.~\ref{results-sun}).

\subsection{Heterogeneous Processes and Galactic Cosmic Rays}
\label{model-detail-het}
Heterogeneous processes which occur on the stratospheric sulfate aerosol (SSA) layer and on polar stratospheric clouds (PSCs) are included as described in \citet{cdj94}.  SSAs are primarily liquid droplets of  $\mathrm{H_2SO_4}$ and $\mathrm{H_2O}$.  PSCs comprise two classes; Type 1 is primarily frozen nitric acid trihydrate (NAT), Type 2 is frozen $\mathrm{H_2O}$.  The formation temperatures of PSCs are low and they occur most frequently in the polar regions, while SSAs are more uniformly distributed (with latitude) in the stratosphere.  Heterogeneous processes occurring on the surfaces of these particulates (especially PSCs) plays an important role in the depletion of ozone and are partly responsible for the preferential depletion in the south polar region where formation of PSCs is more probable.  The formation of PSCs depends on the concentrations of $\mathrm{NO_y}$ (particularly $\mathrm{HNO_3}$) and $\mathrm{H_2O}$.  

In the model, computed concentrations of $\mathrm{NO_y}$ and $\mathrm{H_2O}$ are combined with climatological temperature distributions (obtained from National Meteorological Center data) to calculate: the probability of cloud occurrence; the amount of $\mathrm{H_2O}$ and $\mathrm{HNO_3}$ removed from gas phase in producing clouds; and the surface area density of clouds.  Surface area density is important in computing rates of reactions which occur on the particulates.  It is important to note that this is not a microphysical model of PSCs.  While the computed characteristics of PSCs depends on several assumptions, the results are reasonably insensitive to large changes in the assumptions \citep{cdj94}.  

The effects of galactic cosmic rays (GCRs) are also included in the model.  Ionization rate profiles produced by GCRs were calculated by \citet{nic75} for both solar maximum and minimum.  Rates in the model for a given year are calculated using the yearly averaged monthly sunspot number by linear interpolation of Nicolet's rates.


\clearpage

\clearpage

\clearpage

\clearpage

\clearpage

\clearpage

\clearpage

\clearpage
\begin{deluxetable}{c|cccc}
\tablecaption{Maximum globally averaged $\mathrm{O_3}$ percent change, for $100~\mathrm{kJ/m^2}$ bursts.
              \label{tbl:1}}
\tablewidth{0pt}
\tablehead{
\colhead{Latitude (deg)} & \colhead{March} & \colhead{June} & \colhead{September} & \colhead{December}}
\startdata
+90 & -26 & -29 & -32 & -31 \\
+45 & -30 & -32 & -33 & -31 \\
  0 & -36 & -37 & -38 & -36 \\
-45 & -32 & -31 & -31 & -31 \\
-90 & -31 & -30 & -26 & -36 \\
\enddata
\end{deluxetable}

\clearpage
\begin{deluxetable}{c|cccc}
\tablecaption{Maximum localized $\mathrm{O_3}$ percent change, for $100~\mathrm{kJ/m^2}$ bursts.
              \label{tbl:2}}
\tablewidth{0pt}
\tablehead{
\colhead{Latitude (deg)} & \colhead{March} & \colhead{June} & \colhead{September} & \colhead{December}}
\startdata
+90 & -60 & -66 & -74 & -70 \\
+45 & -56 & -63 & -72 & -66 \\
  0 & -55 & -55 & -60 & -55 \\
-45 & -67 & -63 & -55 & -66 \\
-90 & -71 & -66 & -59 & -62 \\
\enddata
\end{deluxetable}

\clearpage

\clearpage

\clearpage

\clearpage
\begin{deluxetable}{ccc}
\tablecaption{Maximum globally averaged $\mathrm{O_3}$ percent change, every 2 hours, for $100~\mathrm{kJ/m^2}$, March, equator burst case.
              \label{tbl:3}}
\tablewidth{0pt}
\tablehead{
\colhead{Time of Day} & \colhead{Global Average} & \colhead{Localized}}
\startdata
Noon & -36 & -55 \\
2pm & -37 & -55 \\
4pm & -37 & -55 \\
6pm & -36 & -55 \\
8pm & -40 & -57 \\
10pm & -40 & -57 \\
Midnight & -40 & -57 \\
2am & -40 & -57 \\
4am & -40 & -57 \\
6am & -37 & -55 \\
8am & -37 & -55 \\
10am & -37 & -55 \\
\enddata
\end{deluxetable}

\clearpage

\clearpage

\clearpage

\clearpage

\clearpage

\clearpage

\clearpage
\begin{deluxetable}{ccc}
\tablecaption{Maximum $\mathrm{O_3}$ percent changes for $10~\mathrm{kJ/m^2}$ cases.
              \label{tbl:4}}
\tablewidth{0pt}
\tablehead{
\colhead{Burst Case} & \colhead{Global Average} & \colhead{Localized}}
\startdata
March, Equator & -16 & -28 \\
September, $+45^\circ$ & -13 & -38 \\
\enddata
\end{deluxetable}

\clearpage

\clearpage

\clearpage

\clearpage

\clearpage
\begin{deluxetable}{c|cccc}
\tablecaption{Annual global average nitrate deposition ($10^{-3}~\mathrm{g/m^2}$) above baseline in year 3 (post-burst), for $100~\mathrm{kJ/m^2}$ bursts. \label{tbl:no3}}
\tablewidth{0pt}
\tablehead{
\colhead{Latitude (deg)} & \colhead{March} & \colhead{June} & \colhead{September} & \colhead{December}}
\startdata
+90 & 5.19 & 6.52 & 10.9 & 8.98 \\
+45 & 4.45 & 5.62 & 9.36 & 6.87 \\
  0 & 3.52 & 3.37 & 5.67 & 4.17 \\
-45 & 3.83 & 2.72 & 2.84 & 2.72 \\
-90 & 4.40 & 2.95 & 2.49 & 2.68 \\
\enddata
\end{deluxetable}

\clearpage

\clearpage
\begin{deluxetable}{cccc}
\tablecaption{Species Included in the Model \label{tbl:species}}
\tablewidth{0pt}
\tablehead{\colhead{Number} & \colhead{Species} & \colhead{Number} & \colhead{Species}}
\startdata
1 & $\mathrm{O(^3P)}$   & 34 & $\mathrm{HO_2NO_2}$ \\
2 & $\mathrm{O(^1D)}$   & 35 & $\mathrm{Cl_y}$ \\
3 & $\mathrm{O_2}$      & 36 & $\mathrm{CFCl_3}$ \\ 
4 & $\mathrm{O_3}$      & 37 & $\mathrm{CF_2Cl_2}$ \\
5 & $\mathrm{NO}$       & 38 & $\mathrm{CCl_4}$  \\
6 & $\mathrm{NO_2}$     & 39 & $\mathrm{CH_3Cl}$ \\
7 & $\mathrm{NO_3}$     & 40 & liquid $\mathrm{H_2O}$ \\
8 & $\mathrm{N_2O_5}$   & 41 & $\mathrm{O_x}$ \\
9 & $\mathrm{N}$        & 42 & $\mathrm{CH_3CCl_3}$ \\
10 & $\mathrm{N_2}$     & 43 & $\mathrm{O_2(^1\Delta)}$ \\
11 & $\mathrm{HNO_3}$   & 44 & $\mathrm{N(^2D)}$ \\
12 & $\mathrm{N_2O}$    & 45 & $\mathrm{CO_2}$  \\
13 & $\mathrm{H}$       & 46 & $\mathrm{BrO}$ \\
14 & $\mathrm{OH}$      & 47 & $\mathrm{Br}$  \\
15 & $\mathrm{HO_2}$    & 48 & $\mathrm{HBr}$ \\
16 & $\mathrm{H_2O}$    & 49 & $\mathrm{BrONO_2}$ \\
17 & $\mathrm{H_2O_2}$  & 50 & $\mathrm{Br_x}$ \\
18 & $\mathrm{HCl}$     & 51 & $\mathrm{CH_3Br}$ \\
19 & $\mathrm{H_2}$     & 52 & $\mathrm{CBrF_3}$  \\
20 & $\mathrm{CH_4}$    & 53 & $\mathrm{CBrClF_2}$ \\
21 & $\mathrm{CO}$      & 54 & $\mathrm{CHClF_2}$ \\
22 & $\mathrm{CH_3}$    & 55 & $\mathrm{C_2Cl_3F_3}$ \\
23 & $\mathrm{CH_3O_2}$ & 56 & $\mathrm{C_2Cl_2F_4}$ \\
24 & $\mathrm{CH_3O}$   & 57 & $\mathrm{C_2ClF_5}$ \\
25 & $\mathrm{H_2CO}$   & 58 & $\mathrm{HF}$    \\
26 & $\mathrm{HCO}$     & 59 & $\mathrm{CClFO}$  \\
27 & $\mathrm{HOCl}$    & 60 & $\mathrm{CF_2O}$   \\
28 & $\mathrm{CH_3OOH}$ & 61 & $\mathrm{BrCl}$  \\
29 & $\mathrm{Cl}$      & 62 & $\mathrm{Cl_2O_2}$  \\
30 & $\mathrm{ClO}$     & 63 & $\mathrm{ClO_x}$  \\
31 & $\mathrm{HCl}$     & 64 & $\mathrm{ClONO}$  \\
32 & $\mathrm{ClONO_2}$ & 65 & $\mathrm{Cl_2}$  \\
33 & $\mathrm{NO_y}$    &    &                    \\
\enddata
\end{deluxetable}

\clearpage
\begin{deluxetable}{cccc}
\tablecaption{Transported Species and Families \label{tbl:trans-spc}}
\tablewidth{0pt}
\tablehead{\colhead{Number} & \colhead{Species or Family} & \colhead{Number} & \colhead{Species or Family}}
\startdata
1 & $\mathrm{O_x}$   & 20 & $\mathrm{C_2ClF_5}$ \\
2 & $\mathrm{NO_y}$   & 21 & $\mathrm{CBrClF_2}$ \\
3 & $\mathrm{Cl_y}$      & 22 & $\mathrm{CBrF_3}$ \\ 
4 & $\mathrm{N_2O}$      & 23 & $\mathrm{HF}$ \\
5 & $\mathrm{CFCl_3}$       & 24 & $\mathrm{CClFO}$  \\
6 & $\mathrm{CF_2Cl_2}$     & 25 & $\mathrm{CF_2O}$ \\
7 & $\mathrm{CCl_4}$     & 26 & $\mathrm{ClONO_2}$ \\
8 & $\mathrm{CH_3Cl}$   & 27 & $\mathrm{N_2O_5}$ \\
9 & $\mathrm{CH_4}$        & 28 & $\mathrm{N_2O}$ \\
10 & $\mathrm{H_2}$     & 29 & $\mathrm{HNO_3 ~(solid)}$ \\
11 & $\mathrm{CO}$   & 30 & $\mathrm{H_2O ~(solid)}$ \\
12 & $\mathrm{CH_x}$    & 31 & $\mathrm{CO_2}$  \\
13 & $\mathrm{CH_3CCl_3}$       & 32 & $\mathrm{CH_3CCl_2F}$ \\
14 & $\mathrm{HNO_3}$         & 33 & $\mathrm{CH_3CClF_2}$ \\
15 & $\mathrm{Br_y}$        & 34 & $\mathrm{CHCl_2CF_3}$ \\
16 & $\mathrm{CH_3Br}$       & 35 & $\mathrm{C_2Br_2F_4}$ \\
17 & $\mathrm{CHClF_2}$      & 36 & $\mathrm{H_2O_2}$ \\
18 & $\mathrm{C_2Cl_3F_3}$   & 37 & $\mathrm{HO_x}$ \\
19 & $\mathrm{C_2Cl_2F_4}$   
\enddata
\tablecomments{Families are those with ``x'' or ``y'' subscript.}
\end{deluxetable}

\clearpage
\begin{deluxetable}{lcl}
\tablecaption{Lower Boundary Conditions for Transported Species \label{tbl:BCs}}
\tablewidth{0pt}
\tablehead{
\colhead{Species} & \colhead{Boundary Condition Type} & \colhead{Value}}
\startdata
$\mathrm{N_2O}$ & mixing ratio & 270 ppbv  \\
$\mathrm{CH_4}$ & mixing ratio & 700 ppbv  \\
$\mathrm{CO_2}$ & mixing ratio & 280 ppmv  \\
$\mathrm{CO}$ & mixing ratio & 100 ppbv  \\
$\mathrm{CH_3Br}$ & mixing ratio & 7.0 pptv  \\
$\mathrm{CH_3Cl}$ & mixing ratio & 483 pptv  \\
$\mathrm{H_2}$ & mixing ratio & 500 ppbv  \\
$\mathrm{NO_y - HNO_3}$ & mixing ratio & 105 pptv  \\
$\mathrm{CCl_4}$ & mixing ratio & 0.0  \\
$\mathrm{CH_3CCl_3}$ & mixing ratio & 0.0  \\
$\mathrm{Cl_y}$ & flux & 0.0 $\mathrm{cm^{-2} s^{-1}}$ \\
$\mathrm{Br_y}$ & flux & 0.0 $\mathrm{cm^{-2} s^{-1}}$ \\
$\mathrm{HF}$ & flux & 0.0 $\mathrm{cm^{-2} s^{-1}}$ \\
$\mathrm{CClFO}$ & flux & 0.0 $\mathrm{cm^{-2} s^{-1}}$ \\
$\mathrm{CF_2O}$ & flux & 0.0 $\mathrm{cm^{-2} s^{-1}}$ \\
$\mathrm{CH_3OOH}$ & flux & 0.0 $\mathrm{cm^{-2} s^{-1}}$ \\
$\mathrm{O_x}$ & deposition velocity & 0.1 $\mathrm{cm~s^{-1}}$ \\
\enddata
\end{deluxetable}

\clearpage
\begin{deluxetable}{ll}
\tablecaption{Binary Reactions and Rate Constants \label{tbl:rxns1}}
\tablewidth{0pt}
\tablehead{
\colhead{Reaction} & \colhead{Rate Constant}}
\startdata
  $\mathrm{O + O_3 \rightarrow O_2 + O_2}$ & $\mathrm{k_1 = 8.0(-12)~exp(-2060/T)}$   \\
  $\mathrm{OH + O_3 \rightarrow HO_2 + O_2}$ & $\mathrm{k_2 = 1.5(-12)~exp(-880/T)}$   \\
  $\mathrm{HO_2 + O_3 \rightarrow OH + 2O_2}$ & $\mathrm{k_3 = 2.0(-14)~exp(-680/T)}$  \\
  $\mathrm{ClO + HO_2 \rightarrow HOCl + O_2}$ & $\mathrm{k_4 = 4.8(-13)~exp(700/T)}$  \\
  $\mathrm{Cl + H_2O_2 \rightarrow HCl + HO_2}$ & $\mathrm{k_5 = 1.1(-11)~exp(-980/T)}$  \\
  $\mathrm{O(^1D) + M \rightarrow O(^3P) + M}$  & $\mathrm{k_6 = 0.78[1.8(-11)~exp(110/T)]}$ \\ 
                                                  & $\mathrm{+0.21[3.2(-11)~exp(70/T)]}$ \\
  $\mathrm{NO + O_3 \rightarrow NO_2 + O_2}$  & $\mathrm{k_7 = 3.0(-12)~exp(-1500/T)}$  \\
  $\mathrm{NO_2 + O_3 \rightarrow NO_3 + O_2}$ & $\mathrm{k_8 = 1.2(-13)~exp(-2450/T)}$  \\
  $\mathrm{H + O_3 \rightarrow OH + O_2}$ & $\mathrm{k_9 = 1.4(-10)~exp(-470/T)}$  \\
  $\mathrm{OH + ClONO_2 \rightarrow HOCl + NO_3}$ & $\mathrm{k_{10} = 1.2(-12)~exp(-330/T)}$  \\
  $\mathrm{CH_4 + OH \rightarrow CH_3 + H_2O}$ & $\mathrm{k_{11} = 2.45(-12)~exp(-1775/T)}$  \\
  $\mathrm{CH_3O_2 + NO \rightarrow CH_3O + NO_2}$ & $\mathrm{k_{12} = 3.0(-12)~exp(280/T)}$  \\
  $\mathrm{OH + CH_3Cl \rightarrow H_2O + CH_2Cl}$ & $\mathrm{k_{13} = 4.0(-12)~exp(-1400/T)}$  \\
  $\mathrm{CH_3O + O_2 \rightarrow CH_2O + HO_2}$ & $\mathrm{k_{14} = 3.9(-14)~exp(-900/T)}$  \\
  $\mathrm{HO_2 + HO_2 \rightarrow H_2O_2 + O_2}$ & $\mathrm{k_{15} = 2.3(-13)~exp(600/T)}$  \\
  $\mathrm{N + O_2 \rightarrow NO + O}$ & $\mathrm{k_{16} = 1.5(-11)~exp(-3600/T)}$  \\
  $\mathrm{CH_2O + O \rightarrow HCO + OH}$ & $\mathrm{k_{17} = 3.4(-11)~exp(-1600/T)}$  \\
  $\mathrm{CH_3O_2 + HO_2 \rightarrow CH_3OOH + O_2}$ & $\mathrm{k_{18} = 3.8(-13)~exp(800/T)}$ \\
  $\mathrm{Cl + H_2 \rightarrow HCl + H}$ & $\mathrm{k_{19} = 3.7(-11)~exp(-2300/T)}$  \\
  $\mathrm{Cl + O_3 \rightarrow ClO + O_2}$ & $\mathrm{k_{20} = 2.3(-11)~exp(-200/T)}$  \\
  $\mathrm{ClO + O \rightarrow Cl + O_2}$ & $\mathrm{k_{21} = 3.0(-11)~exp(-70/T)}$  \\
  $\mathrm{Cl + CH_4 \rightarrow HCl + CH_3}$ & $\mathrm{k_{22} = 9.6(-12)~exp(-1360/T)}$  \\
  $\mathrm{HCl + OH \rightarrow Cl + H_2O}$ & $\mathrm{k_{23} = 2.6(-12)~exp(-350/T)}$  \\
  $\mathrm{ClO + NO \rightarrow Cl + NO_2}$ & $\mathrm{k_{24} = 6.4(-12)~exp(290/T)}$  \\
  $\mathrm{OH + H_2O_2 \rightarrow H_2O + HO_2}$ & $\mathrm{k_{25} = 2.9(-12)~exp(-160/T)}$  \\
  $\mathrm{H_2 + OH \rightarrow H_2O + H}$ & $\mathrm{k_{26} = 5.5(-12)~exp(-2000/T)}$  \\
  $\mathrm{N_2O_5 + M \rightarrow NO_2 + NO_3 + M}$ & $\mathrm{k_{27} = 2.7(-27)~exp(-11000/T)}$ \\
  $\mathrm{O + H_2O_2 \rightarrow OH + HO_2}$ & $\mathrm{k_{28} = 1.4(-12)~exp(-2000/T)}$  \\
  $\mathrm{O + ClONO_2 \rightarrow ClO + NO_3}$ & $\mathrm{k_{29} = 2.9(-12)~exp(-800/T)}$  \\
  $\mathrm{CO + OH \rightarrow CO_2 + H}$ & $\mathrm{k_{30} = 1.5(-13)}$  \\
%
  $\mathrm{HNO_3 + OH \rightarrow NO_3 + H_20}$ & $\mathrm{k_{31} = 7.2(-15)~exp(-785/T)}$ \\
		                                & $\mathrm{+(N_d 1.9(-33)~exp(-725/T)/}$ \\
						& $\mathrm{(1+N_d 1.9(-33)~exp(-725/T)/}$ \\
						& $\mathrm{(4.1(-16)~exp(-1440/T))))}$ \\
  $\mathrm{NO + HO_2 \rightarrow OH + NO_2}$ & $\mathrm{k_{32} = 3.5(-12)~exp(250/T)}$  \\
  $\mathrm{H_2O + O(^1D) \rightarrow OH + OH}$ & $\mathrm{k_{33} = 2.2(-10)}$  \\
  $\mathrm{OH + HO_2 \rightarrow H_2O + O_2}$ & $\mathrm{k_{34} = 4.8(-11)~exp(250/T)}$  \\
  $\mathrm{OH + O \rightarrow H + O_2}$ & $\mathrm{k_{35} = 2.2(-11)~exp(120/T)}$  \\
  $\mathrm{HO_2 + O \rightarrow OH + O_2}$ & $\mathrm{k_{36} = 3.0(-11)~exp(200/T)}$  \\
  $\mathrm{NO_2 + O \rightarrow NO + O_2}$ & $\mathrm{k_{37} = 5.6(-12)~exp(180/T)}$  \\
  $\mathrm{N_2O + O(^1D) \rightarrow NO + NO}$ & $\mathrm{k_{38} = 6.7(-11)}$ \\
  $\mathrm{N + NO \rightarrow N_2 + O}$ & $\mathrm{k_{39} = 2.1(-11)~exp(100/T)}$  \\
  $\mathrm{H_2 + O(^1D) \rightarrow OH + H}$ & $\mathrm{k_{40} = 1.1(-10)}$ \\
  $\mathrm{CH_4 + O(^1D) \rightarrow CH_3 + OH}$ & $\mathrm{k_{41} = 1.12(-10)}$ \\
  $\mathrm{H_2CO + OH \rightarrow H_2O + HCO}$ & $\mathrm{k_{42} = 1.0(-11)}$ \\
  $\mathrm{HCO + O_2 \rightarrow CO + HO_2}$ & $\mathrm{k_{43} = 3.5(-12)~exp(140/T)}$  \\
  $\mathrm{Cl + HO_2 \rightarrow HCl + O_2}$ & $\mathrm{k_{44} = 1.8(-11)~exp(170/T)}$  \\
  $\mathrm{CCl_4 + O(^1D) \rightarrow 4Cl + Prods}$ & $\mathrm{k_{45} = 3.3(-10)}$  \\
  $\mathrm{OH + HO_2NO_2 \rightarrow H_2O + O_2 + NO_2}$ & $\mathrm{k_{46} = 1.3(-12)~exp(380/T)}$ \\
  $\mathrm{CH_4 + O(^1D) \rightarrow H_2 + H_2CO}$ & $\mathrm{k_{47} = 8.0(-12)}$  \\
  $\mathrm{OH + CH_3OOH \rightarrow H_2O + CH_3O_2}$ & $\mathrm{k_{48} = 3.8(-12)~exp(200/T)}$  \\
  $\mathrm{OH + OH \rightarrow H_2O + O}$ & $\mathrm{k_{49} = 4.2(-12)~exp(-240/T)}$  \\
  $\mathrm{ClO + OH \rightarrow Cl + HO_2}$ & $\mathrm{k_{50} = 7.4(-12)~exp(270/T)}$  \\
  $\mathrm{ClO + OH \rightarrow HCl + O_2}$ & $\mathrm{k_{51} = 3.2(-13)~exp(320/T)}$  \\
  $\mathrm{HOCl + OH \rightarrow H_2O + ClO}$ & $\mathrm{k_{52} = 3.0(-12)~exp(-500/T)}$  \\
  $\mathrm{Cl + H_2CO \rightarrow HCl + HCO}$ & $\mathrm{k_{53} = 8.1(-11)~exp(-30/T)}$  \\
  $\mathrm{HO_2 + HO_2 + M \rightarrow H_2O_2 + O_2 + M}$ & $\mathrm{k_{54} = 1.7(-33)~exp(1000/T)}$ \\
  $\mathrm{CFClO  +  O(^1D)  \rightarrow  Prods}$ & $\mathrm{k_{55} = 1.9(-10)}$ \\
  $\mathrm{CF_2O  +  O(^1D)   \rightarrow  Prods}$ & $\mathrm{k_{56} = 7.4(-11)}$ \\
  $\mathrm{Cl + HO_2 \rightarrow OH + ClO}$ & $\mathrm{k_{57} = 4.1(-11)~exp(-450/T)}$  \\
  $\mathrm{N + OH \rightarrow NO + H}$ & $\mathrm{k_{58} = 5.0(-11)}$  \\
  $\mathrm{BrO + NO \rightarrow NO_2 + Br}$ & $\mathrm{k_{59} = 8.8(-12)~exp(260/T)}$  \\
  $\mathrm{HO_2NO_2 + M \rightarrow HO_2 + NO_2 + M}$ & $\mathrm{k_{60} = 2.1(-27)~exp(-10900/T)}$  \\
  $\mathrm{H + HO_2 \rightarrow H_2 + O_2}$ & $\mathrm{k_{61} = 1.06(-11)}$  \\
  $\mathrm{H + HO_2 \rightarrow H_2O + O}$ & $\mathrm{k_{62} = 1.60(-12)}$  \\
  $\mathrm{H + HO_2 \rightarrow OH + OH}$ & $\mathrm{k_{63} = 6.88(-11)}$  \\
  $\mathrm{NO + NO_3 \rightarrow NO_2 + NO_2}$ & $\mathrm{k_{64} = 1.5(-11)~exp(170/T)}$  \\
  $\mathrm{OH + CH_3CCl_3 \rightarrow 3(Cl)}$ & $\mathrm{k_{65} = 1.8(-12)~exp(-1550/T)}$  \\
  $\mathrm{N(^2D) + O_2 \rightarrow NO + O}$ & $\mathrm{k_{66} = 5.0(-11)}$  \\
  $\mathrm{N_2O + O(^1D) \rightarrow N_2 + O_2}$ & $\mathrm{k_{67} = 4.9(-11)}$  \\
  $\mathrm{O(^1D) + CF_2Cl_2 \rightarrow ClO + Cl + Prods}$ & $\mathrm{k_{68} = 1.4(-10)}$  \\
  $\mathrm{N + NO_2 \rightarrow N_2O + O}$ & $\mathrm{k_{69} = 5.8(-12)~exp(220/T)}$  \\
  $\mathrm{O(^1D) + CFCl_3 \rightarrow ClO + 2Cl + Prods}$ & $\mathrm{k_{70} = 2.3(-10)}$  \\
  $\mathrm{O(^1D) + DECAY \rightarrow O + h\nu}$ & $\mathrm{k_{71} = 6.8(-09)}$  \\
  $\mathrm{O_2(^1\Delta) + DECAY \rightarrow O_2 + h\nu}$ & $\mathrm{k_{72} = 2.6(-10)}$  \\
  $\mathrm{N(^2D) + O \rightarrow N + O}$ & $\mathrm{k_{73} = 1.4(-10)}$  \\
  $\mathrm{O_2(^1\Delta) + O_2 \rightarrow O_2 + O_2}$ & $\mathrm{k_{74} = 3.6(-18)~exp(-220/T)}$  \\
  $\mathrm{O_2(^1\Delta) + O_3 \rightarrow O_2 + O_2 + O}$ & $\mathrm{k_{75} = 5.2(-11)~exp(-2840/T)}$ \\
  $\mathrm{O_2(^1\Delta) + NO \rightarrow O_2 + NO}$ & $\mathrm{k_{76} = 4.5(-17)}$ \\
  $\mathrm{O + O + M \rightarrow O_2 + M}$ & $\mathrm{k_{77} = 1.4(-33)~exp(408/T)}$  \\
  $\mathrm{Br + O_3 \rightarrow BrO + O_2}$ & $\mathrm{k_{78} = 1.7(-11)~exp(-800/T)}$  \\
  $\mathrm{Br + HO_2 \rightarrow HBr + O_2}$ & $\mathrm{k_{79} = 1.5(-11)~exp(-600/T)}$  \\
  $\mathrm{BrO + ClO  \rightarrow  Br  +  ClOO}$ & $\mathrm{k_{80} = 2.3(-12)~exp(260/T)}$  \\
  $\mathrm{BrO + BrO \rightarrow Br + Br + O_2}$ & $\mathrm{k_{81} = 1.5(-12)~exp(230/T)}$  \\
  $\mathrm{OH + HBr \rightarrow H_2O + Br}$ & $\mathrm{k_{82} = 1.1(-11)}$  \\
  $\mathrm{CH_3Br + OH \rightarrow Br + Prods}$ & $\mathrm{k_{83} = 4.0(-12)~exp(-1470/T)}$  \\
  $\mathrm{CHClF_2 + OH \rightarrow Cl + 2F + Prods}$ & $\mathrm{k_{84} = 1.0(-12)~exp(-1600/T)}$ \\
  $\mathrm{C_2Cl_3F_3 + O(^1D) \rightarrow 3Cl + 3F + Prods}$ & $\mathrm{k_{85} = 2.00(-10)}$  \\
  $\mathrm{C_2Cl_2F_4 + O(^1D) \rightarrow 2Cl + 4F + Prods}$ & $\mathrm{k_{86} = 1.00(-10)}$  \\
  $\mathrm{C_2ClF_5 + O(^1D) \rightarrow Cl + 5F + Prods}$ & $\mathrm{k_{87} = 5.00(-11)}$  \\
  $\mathrm{BrO + ClO  \rightarrow  Br  +  OClO}$ & $\mathrm{k_{88} = 9.5(-13)~exp(550/T)}$  \\
  $\mathrm{BrO + ClO  \rightarrow  BrCl  +  O_2}$ & $\mathrm{k_{89} = 4.1(-13)~exp(290/T)}$  \\
  $\mathrm{Cl_2O_2  +  M  \rightarrow  ClO  +  ClO}$ & $\mathrm{k_{90} = 1.3(-27)~exp(-8744/T)}$  \\
  $\mathrm{BrO + O \rightarrow Br + O_2}$ & $\mathrm{k_{91} = 1.9(-11)~exp(230/T)}$  \\
  $\mathrm{BrO + HO_2 \rightarrow HOBr + O_2}$ & $\mathrm{k_{92} = 3.4(-12)~exp(540/T)}$  \\
  $\mathrm{Br + CH_2O \rightarrow HBr + CHO}$ & $\mathrm{k_{93} = 1.7(-11)~exp(-800/T)}$  \\
  $\mathrm{CH_4 + O(^1D) \rightarrow H + CH_3O}$ & $\mathrm{k_{94} = 3.0(-11)}$  \\
  $\mathrm{O(^1D) + CClBrF_2 \rightarrow Prods}$ & $\mathrm{k_{95} = 1.5(-10)}$  \\
  $\mathrm{O(^1D) + CBrF_3 \rightarrow Prods}$ & $\mathrm{k_{96} = 1.0(-10)}$  \\
  $\mathrm{O(^1D) + CH_3CCl_2F \rightarrow Prods}$ & $\mathrm{k_{97} = 2.6(-10)}$  \\
  $\mathrm{OH + CH_3CCl_2F \rightarrow Prods + H_2O}$ & $\mathrm{k_{98} = 1.7(-12)~exp(-1700/T)}$ \\
  $\mathrm{Cl + CH_3CCl_2F \rightarrow Prods + HCl}$ & $\mathrm{k_{99} = 1.8(-12)~exp(-2000/T)}$  \\
  $\mathrm{O(^1D) + CH_3CF_2Cl \rightarrow Prods}$ & $\mathrm{k_{100} = 2.2(-10)}$  \\
  $\mathrm{OH + CH_3CF_2Cl \rightarrow Prods + H_2O}$ & $\mathrm{k_{101} = 1.3(-12)~exp(-1800/T)}$ \\
  $\mathrm{Cl + CH_3CF_2Cl \rightarrow Prods + HCl}$ & $\mathrm{k_{102} = 1.4(-12)~exp(-2420/T)}$ \\
  $\mathrm{O(^1D) + CHCl_2CF_3 \rightarrow Prods}$ & $\mathrm{k_{103} = 2.0(-10)}$ \\
  $\mathrm{OH + CHCl_2CF_3 \rightarrow Prods + H_2O}$ & $\mathrm{k_{104} = 7.0(-13)~exp(-900/T)}$ \\
  $\mathrm{Cl + CHCl_2CF_3 \rightarrow Prods + HCl}$ & $\mathrm{k_{105} = 4.4(-12)~exp(-1750/T)}$ \\
  $\mathrm{O(^1D) + C_2Br_2F_4 \rightarrow Prods}$ & $\mathrm{k_{106} = 1.6(-10)}$ \\
  $\mathrm{O(^1D) + CH_3Br  \rightarrow  Br + CH_3O}$ & $\mathrm{k_{107} = 1.8(-10)}$ \\
  $\mathrm{O(^1D) + CHClF_2  \rightarrow  Cl + CF_2O + H}$ & $\mathrm{k_{108} = 1.0(-10)}$ \\
  $\mathrm{O + CH_3  \rightarrow  CH_2O +  H}$ & $\mathrm{k_{109} = 1.1(-10)}$ \\
  $\mathrm{O_3 + CH_3  \rightarrow  CH_3O + O_2}$ & $\mathrm{k_{110} = 5.4(-12)~exp(-220/T)}$ \\
  $\mathrm{CH_3O_2 + CH_3O_2  \rightarrow  2CH_3O + O_2}$ & $\mathrm{k_{111} = 2.5(-13)~exp(190/T)}$ \\
  $\mathrm{CFCl_3 + OH  \rightarrow  3Cl + HF + CO}$ & $\mathrm{k_{112} = 1.0(-12)~exp(-3700/T)}$ \\
  $\mathrm{CF_2Cl_2 + OH  \rightarrow Cl + HF + CClFO}$ & $\mathrm{k_{113} = 1.0(-12)~exp(-3600/T)}$ \\
  $\mathrm{CH_3Cl + Cl  \rightarrow  2Cl + CH_3}$ & $\mathrm{k_{114} = 3.2(-11)~exp(-1250/T)}$  \\
\enddata
\tablecomments{Read ``Prods'' as ``Products.'' Read $1.0(-10)$ as $1.0 \times 10^{-10}$.
$\mathrm{M = N_2, O_2}$ (arbitrary third body). $\mathrm{N_d}$ = total number density at a particular grid point.}
\end{deluxetable}

\clearpage
\begin{deluxetable}{lcccc}
\tablecaption{Tertiary Reactions and Rate Constants \label{tbl:rxns2}}
\tablewidth{0pt}
\tablehead{
\colhead{Reaction} & \colhead{$k_0^{300}$} & \colhead{n} & \colhead{$k_{\inf}^{300}$} & \colhead{m} }
\startdata
  $\mathrm{O + O_2 + M \rightarrow O_3 + M}$ & $6.0(-34)$ & $2.3$ & $0.0$ & $0.0$ \\ 
  $\mathrm{H + O_2 + M \rightarrow HO_2 + M}$ & $5.7(-32)$ & $1.6$ & $7.5(-11)$ & $0.0$ \\ 
  $\mathrm{OH + OH + M \rightarrow H_2O_2 + M}$ & $6.2(-31)$ & $1.0$ & $2.6(-11)$ & $0.0$ \\ 
  $\mathrm{OH + NO_2 + M \rightarrow HNO_3 + M}$ & $2.5(-30)$ & $4.4$ & $1.6(-11)$ & $1.7$ \\ 
  $\mathrm{ClO + NO_2 + M \rightarrow ClONO_2 + M}$ & $1.8(-31)$ & $3.4$ & $1.5(-11)$ & $1.9$ \\ 
  $\mathrm{HO_2 + NO_2 + M \rightarrow HO_2NO_2 + M}$ & $1.8(-31)$ & $3.2$ & $4.7(-12)$ & $1.4$ \\ 
  $\mathrm{NO_2 + O + M \rightarrow NO_3 + M}$ & $9.0(-32)$ & $2.0$ & $2.2(-11)$ & $0.0$  \\ 
  $\mathrm{NO_2 + NO_3 + M \rightarrow N_2O_5 + M}$ & $2.2(-30)$ & $3.9$ & $1.5(-12)$ & $0.7$ \\ 
  $\mathrm{CH_3 + O_2 + M \rightarrow CH_3O_2 + M}$ & $4.5(-31)$ & $3.0$ & $1.8(-12)$ & $1.7$ \\ 
  $\mathrm{NO + O + M \rightarrow NO_2 + M}$ & $9.0(-32)$ & $1.5$ & $3.0(-11)$ & $0.0$ \\ 
  $\mathrm{O(^1D) + N_2 + M \rightarrow N_2O + M}$ & $3.5(-37)$ & $0.6$ & $0.0$ &  $0.0$ \\ 
  $\mathrm{BrO + NO_2 + M \rightarrow BrONO_2 + M}$ & $5.2(-31)$ & $3.2$ & $6.9(-12)$ &  $2.9$ \\ 
  $\mathrm{ClO + ClO + M \rightarrow Cl_2O_2 + M}$ & $2.2(-32)$ & $3.1$ & $3.5(-12)$ & $1.0$ \\ 
  $\mathrm{CO + O + M \rightarrow CO_2 + M}$ & $2.0(-37)$ & $0.0$ & $0.0$ & $0.0$ \\ 
\enddata
\tablecomments{
Read $1.0(-10)$ as $1.0 \times 10^{-10}$.  $\mathrm{M = N_2, O_2}$ (arbitrary third body)\\
Rate constants are given by: 
$$k = [k_0(T)]/[1+k_0(T)[M]/k_\infty(T)]0.6^{\{1+[log_{10}(k_0(T)[M]/k_\infty(T))]^2\}^{-1}}$$ \\
Low pressure limit: $k_0(T) = k_0^{300}(T/300)^{-n} ~(\mathrm{cm^6 ~s^{-1}})$ \\
High pressure limit: $k_\infty(T) = k_\infty^{300}(T/300)^{-m} ~(\mathrm{cm^3 ~s^{-1}})$ \\
}
\end{deluxetable}

\clearpage
\begin{deluxetable}{l}
\tablecaption{Photodissociations \label{tbl:rxns3}}
\tablewidth{0pt}
\tablehead{\colhead{Reaction}}
\startdata
  $\mathrm{O_2   \rightarrow O + O }$  \\ 
  $\mathrm{O_3	\rightarrow O_2(^1\Delta) + O(^1D) }$  \\ 
  $\mathrm{O_3	\rightarrow O_2 + O }$  \\ 
  $\mathrm{H_2O	\rightarrow H + OH }$  \\ 
  $\mathrm{NO_3	\rightarrow NO_2 + O }$  \\ 
  $\mathrm{HNO_3	\rightarrow OH + NO_2 }$  \\ 
  $\mathrm{NO_2	\rightarrow NO + O }$  \\ 
  $\mathrm{H_2O_2	\rightarrow OH + OH }$  \\ 
  $\mathrm{N_2O_5	\rightarrow NO_2 + NO_3 }$  \\ 
  $\mathrm{H_2CO	\rightarrow HCO + H }$  \\ 
  $\mathrm{H_2CO	\rightarrow H_2 + CO }$  \\ 
  $\mathrm{CO_2	\rightarrow CO + O  }$  \\ 
  $\mathrm{CH_3OOH	\rightarrow CH_3O + OH  }$  \\ 
  $\mathrm{N_2O	\rightarrow N_2 + O  }$  \\ 
  $\mathrm{ClONO_2	\rightarrow Cl + NO_3 }$  \\ 
  $\mathrm{NO	\rightarrow N + O   }$  \\ 
  $\mathrm{NO_3	\rightarrow NO + O_2  }$  \\ 
  $\mathrm{HCl	\rightarrow H + Cl  }$  \\ 
  $\mathrm{CCl_4	\rightarrow 4Cl + Fragment  }$  \\ 
  $\mathrm{CH_3Cl	\rightarrow CH_3 + Cl  }$  \\ 
  $\mathrm{CFCl_3	\rightarrow 3Cl + Fragment }$  \\ 
  $\mathrm{CF_2Cl_2	\rightarrow 2Cl + Fragment }$  \\ 
  $\mathrm{HOCl	\rightarrow OH + Cl   }$  \\ 
  $\mathrm{HO_2NO_2	\rightarrow OH + NO_3   }$  \\ 
  $\mathrm{H_2O	\rightarrow H_2 + O(^1D)  }$  \\ 
  $\mathrm{CH_3CCl_3	\rightarrow 3Cl + Fragment  }$  \\ 
  $\mathrm{BrO	\rightarrow Br + O   }$  \\ 
  $\mathrm{BrONO_2	\rightarrow Br + NO_3  }$  \\ 
  $\mathrm{CH_3Br	\rightarrow CH_3 + Br  }$  \\ 
  $\mathrm{CF_3Br	\rightarrow Br + 3F + Fragment }$  \\ 
  $\mathrm{CF_2ClBr	\rightarrow Br + Cl + 2F + Fragment  }$  \\ 
  $\mathrm{CHClF_2	\rightarrow Cl + 2F + Fragment   }$  \\ 
  $\mathrm{C_2Cl_3F_3	\rightarrow 3Cl + 3F + Fragment   }$  \\ 
  $\mathrm{C_2Cl_2F_4	\rightarrow 2Cl + 4F + Fragment   }$  \\ 
  $\mathrm{C_2ClF_5	\rightarrow Cl + 5F + Fragment  }$  \\ 
  $\mathrm{Cl_2O_2	\rightarrow Cl + OClO  }$  \\ 
  $\mathrm{OClO	\rightarrow Cl + O_2  }$  \\ 
  $\mathrm{Cl_2	\rightarrow Cl + Cl   }$  \\ 
  $\mathrm{ClONO	\rightarrow ClO + NO  }$  \\ 
  $\mathrm{BrCl	\rightarrow Br + Cl  }$  \\ 
  $\mathrm{CO_2     \rightarrow CO + O(^1D)   }$  \\ 
  $\mathrm{HO_2NO_2  \rightarrow HO_2 + NO_2   }$  \\ 
  $\mathrm{CClFO   \rightarrow Cl + F + Fragment  }$  \\ 
  $\mathrm{CF_2O    \rightarrow 2F + Fragment  }$  \\ 
  $\mathrm{N_2O     \rightarrow NO + N  }$  \\ 
  $\mathrm{O_2      \rightarrow O + O(^1D)  }$  \\ 
  $\mathrm{HO_2      \rightarrow O + OH   }$  \\ 
  $\mathrm{CH_3CFCl_2 \rightarrow 2Cl + F + Fragment  }$  \\ 
  $\mathrm{CH_3CF_2Cl \rightarrow Cl + 2F + Fragment   }$  \\ 
  $\mathrm{CF_3CHCl_2 \rightarrow 2Cl + 3F + Fragment    }$  \\ 
  $\mathrm{CF_3CHFCl \rightarrow 4F + Cl + Fragment  }$  \\ 
  $\mathrm{CF_3CF_2CHCl_2 \rightarrow 5F + 2Cl + Fragment  }$  \\ 
  $\mathrm{CF_2ClCF_2CHFCl \rightarrow 5F + 2Cl + Fragment  }$  \\ 
  $\mathrm{CF_2Br_2   \rightarrow 2Br + 2F + Fragment   }$  \\ 
  $\mathrm{CF_2BrCF_2Br \rightarrow 3Br + 3F + Fragment   }$  \\ 
  $\mathrm{CHBr_3    \rightarrow 3Br + Fragment   }$  \\ 
  $\mathrm{CCl_2O    \rightarrow 2Cl + Fragment   }$  \\ 
  $\mathrm{ClONO_2   \rightarrow  ClO + NO_2    }$  \\ 
  $\mathrm{CH_4      \rightarrow CH_3 + H   }$  \\ 
  $\mathrm{CH_4      \rightarrow CH_2 + H_2  }$  \\ 
  $\mathrm{CH_4      \rightarrow CH + H + H_2   }$  \\ 
  $\mathrm{HOBr     \rightarrow HO + Br  }$  \\ 
  $\mathrm{CH_3O_2    \rightarrow  CH_3 + O_2  }$  \\ 
  $\mathrm{BrONO_2   \rightarrow  BrO + NO_2    }$  \\ 
  $\mathrm{CH_3CFCl_2 \rightarrow CO_2 + CClFO + Cl + H + H_2O   }$  \\ 
  $\mathrm{CH_3CF_2Cl \rightarrow CO_2 + CF_2O  + Cl + H + H_2O   }$  \\ 
  $\mathrm{CF_3CHCl_2 \rightarrow CClFO + CF_2O + Cl + H   }$  \\ 
  $\mathrm{C_2F_4Br_2  \rightarrow 2CF_2O + 2Br   }$  \\ 
\enddata
\end{deluxetable}

\clearpage
\begin{deluxetable}{ccc}
\tablecaption{Wavelength Bands and Corresponding Solar Fluxes \label{tbl:bands1}}
\tablewidth{0pt}
\tablehead{
\colhead{Number} & \colhead{Band (nm)} & \colhead{Flux (photons $\mathrm{cm^{-2}~ s^{-1}}$)}}
\startdata
 1 & 121.067 -- 122.067 & $4.006 \times 10^{11}$ \\
 2 & 170.0 -- 172.4 & $1.764 \times 10^{11}$ \\
 3 & 172.4 -- 173.9 & $1.017 \times 10^{11}$ \\ 
 4 & 173.9 -- 175.4 & $1.302 \times 10^{11}$ \\ 
 5 & 175.4 -- 177.0 & $1.722 \times 10^{11}$ \\ 
 6 & 177.0 -- 178.6 & $2.200 \times 10^{11}$ \\ 
 7 & 178.6 -- 180.2 & $2.438 \times 10^{11}$ \\ 
 8 & 180.2 -- 181.8 & $3.215 \times 10^{11}$ \\ 
 9 & 181.8 -- 183.5 & $3.691 \times 10^{11}$ \\ 
10 & 183.5 -- 185.2 & $3.549 \times 10^{11}$ \\ 
11 & 185.2 -- 186.9 & $4.190 \times 10^{11}$ \\ 
12 & 186.9 -- 188.7 & $5.555 \times 10^{11}$ \\ 
13 & 188.7 -- 190.5 & $6.464 \times 10^{11}$ \\ 
14 & 190.5 -- 192.3 & $7.342 \times 10^{11}$ \\ 
15 & 192.3 -- 194.2 & $7.709 \times 10^{11}$ \\ 
16 & 194.2 -- 196.1 & $1.055 \times 10^{12}$ \\ 
17 & 196.1 -- 198.0 & $1.189 \times 10^{12}$ \\ 
18 & 198.0 -- 200.0 & $1.333 \times 10^{12}$ \\ 
19 & 200.0 -- 202.0 & $1.610 \times 10^{12}$ \\ 
20 & 202.0 -- 210.5 & $1.195 \times 10^{13}$ \\ 
21 & 210.5 -- 219.8 & $3.720 \times 10^{13}$ \\ 
22 & 219.8 -- 229.9 & $5.643 \times 10^{13}$ \\ 
23 & 229.9 -- 241.0 & $6.210 \times 10^{13}$ \\ 
24 & 241.0 -- 253.2 & $8.321 \times 10^{13}$ \\ 
25 & 253.2 -- 266.7 & $2.363 \times 10^{14}$ \\ 
26 & 266.7 -- 281.7 & $4.169 \times 10^{14}$ \\ 
27 & 281.7 -- 285.7 & $1.496 \times 10^{14}$ \\ 
28 & 285.7 -- 298.5 & $9.076 \times 10^{14}$ \\ 
29 & 298.5 -- 303.0 & $3.217 \times 10^{14}$ \\ 
30 & 303.0 -- 307.7 & $4.484 \times 10^{14}$ \\ 
31 & 307.7 -- 312.5 & $4.857 \times 10^{14}$ \\ 
32 & 312.5 -- 317.5 & $5.491 \times 10^{14}$ \\ 
33 & 317.5 -- 322.5 & $6.022 \times 10^{14}$ \\ 
34 & 322.5 -- 337.5 & $2.286 \times 10^{15}$ \\ 
35 & 337.5 -- 357.5 & $3.333 \times 10^{15}$ \\ 
36 & 357.5 -- 377.5 & $3.717 \times 10^{15}$ \\ 
37 & 377.5 -- 397.5 & $3.947 \times 10^{15}$ \\ 
38 & 397.5 -- 547.5 & $6.724 \times 10^{16}$ \\ 
39 & 547.5 -- 735.0 & $9.806 \times 10^{16}$ \\ 
\enddata
\end{deluxetable}


\end{document}